\documentclass[10pt, notitlepage, twocolumn, superscriptaddress, nofootinbib, floatfix]{revtex4-1}

\usepackage{amsmath}
\usepackage{mathrsfs}
\usepackage{multirow}
\usepackage{amssymb}
\usepackage{mathtools}
\usepackage{hyperref}
\usepackage[usenames,dvipsnames]{xcolor}
\hypersetup{colorlinks=true, citecolor=Purple, linkcolor=Purple,
urlcolor=Purple}

\usepackage{graphicx}
\usepackage{orcidlink}

\newcommand{\GB}{\mathscr{G}}

\usepackage{tikz}
\usepackage[most]{tcolorbox}
\usepackage{empheq}
\usetikzlibrary{shadows}
\tcbset{
  highlight math style={
    enhanced,
    colframe=blue!40!black,
    colback=blue!10,
    arc=0pt,
    boxrule=1pt,
    drop shadow
  }
}

\begin{document}

\allowdisplaybreaks

\title{Nonperturbative suppression of beyond-General-Relativity effects in quadratic gravity}

\author{Georgios~Antoniou}
\email{georgiosantoniou@tecnico.ulisboa.pt}
\affiliation{CENTRA, Departamento de F\'isica, Instituto Superior T\'ecnico – IST, Universidade de Lisboa – UL, Avenida Rovisco Pais 1, 1049 Lisboa, Portugal}
\affiliation{Dipartimento di Fisica, ``Sapienza'' Universit\'a di Roma, P.A. Moro 5, 00185, Roma, Italy}
\affiliation{Sezione INFN Roma1, P.A. Moro 5, 00185, Roma, Italy}

\author{Leonardo~Gualtieri}
\email{leonardo.gualtieri@unipi.it}
\affiliation{Dipartimento di Fisica,  Universit\`a di Pisa, 56127 Pisa, Italy}
\affiliation{INFN, Sezione di Pisa, Largo B. Pontecorvo 3, 56127 Pisa, Italy}

\author{Paolo~Pani}
\email{paolo.pani@uniroma1.it}
\affiliation{Dipartimento di Fisica, ``Sapienza'' Universit\'a di Roma, P.A. Moro 5, 00185, Roma, Italy}
\affiliation{Sezione INFN Roma1, P.A. Moro 5, 00185, Roma, Italy}

\begin{abstract}
    Quadratic gravity is a well-motivated extension of general relativity~(GR) wherein the Einstein-Hilbert action is augmented by quadratic curvature terms.
   This theory is equivalent to GR in an effective-field-theory framework, while the two theories are different at the non-perturbative level.
    As we have recently shown, black holes in quadratic gravity have a rich linear response, including extra scalar, vector, and tensor quasinormal modes that can be excited in physical processes, even when the stationary solution is the same as in GR.
    Here, by studying the gravitational-wave emission from point particles plunging into a Schwarzschild black hole in quadratic gravity, we show that observable deviations from GR are \emph{exponentially} suppressed in the GR limit.
    This provides a nonperturbative realization of the equivalence between quadratic gravity and GR predicted in the effective-field-theory framework.
\end{abstract}

\maketitle

\section{Introduction}
\label{sec:intro}

One of the most natural and well-motivated avenues for extending general relativity (GR) is through the inclusion of higher-derivative curvature terms in the gravitational action, augmenting the standard Einstein–Hilbert term with operators quadratic or higher in the curvature tensors~\cite{Berti:2015itd}. These corrections can be motivated both from a top-down perspective --~arising, for instance, in the low-energy limit of candidate quantum gravity theories~-- and from a bottom-up effective field theory~(EFT) approach, where all terms compatible with the symmetries of GR are systematically included.

A particularly well-studied class of such modifications is quadratic gravity, where the action includes terms proportional to quadratic contractions of the Riemann tensor, such as $R^2$, $R_{\mu\nu} R^{\mu\nu}$, and $R_{\mu\nu\rho\sigma} R^{\mu\nu\rho\sigma}$. At the linearized level, these terms modify the graviton propagator and introduce additional propagating degrees of freedom: a massive scalar (spin-0) mode and a massive spin-2 mode. While such corrections render the theory formally renormalizable~\cite{Stelle:1976gc}, they also introduce ghost-like instabilities due to the presence of higher-order time derivatives, raising concerns about the theory’s consistency at the fundamental level.

Remarkably, if quadratic gravity is considered as an EFT correction to GR, it is {\it equivalent} to it (as long as matter is not present), since the quadratic terms can be reabsorbed in the Einstein-Hilbert action via field redefinitions at any order in the EFT expansion\,\cite{Burgess:2003jk,Endlich:2017tqa}. Thus, it can be considered as a nontrivial modification to vacuum GR only if treated as a full theory; this, of course, makes the presence of ghost modes more problematic.

Nevertheless, it has been suggested - in different contexts - that quadratic gravity may be viable despite the presence of a ghost~\cite{Donoghue:2019fcb,Donoghue:2021cza,Edelstein:2021jyu,Edelstein:2024jzu}. Thus, its viability as a full theory remains an open issue.
Furthermore, despite the presence of higher-derivative terms and ghost modes, quadratic gravity admits a well-posed initial value formulation in suitable gauges~\cite{Noakes:1983xd}, which allows for the consistent evolution of initial data at the classical level. This has motivated a resurgence of interest in the classical and phenomenological implications of quadratic gravity beyond the strict EFT regime. Recent work has explored black hole solutions~\cite{Lu:2015cqa,Lu:2015psa,Kokkotas:2017zwt,Konoplya:2022iyn}, stability analyses~\cite{East:2023nsk}, perturbative~\cite{Antoniou:2024jku,Knoska:2025jdj,Antoniou:2026swg} and post-Newtonian~\cite{Alves:2025qcx} studies, and numerical simulations~\cite{Held:2023aap,Held:2025ckb}, aiming to understand the full nonlinear dynamics of the theory and its potential observational signatures.

Despite the equivalence between quadratic gravity within an EFT framework and GR in vacuum~\cite{Burgess:2003jk,Endlich:2017tqa},
we have recently shown that the spectrum of black hole quasinormal modes (QNMs) in quadratic gravity has a remarkably rich structure~\cite{Antoniou:2024jku}. In addition to the standard tensorial GR modes, the spectrum contains additional  massive modes associated with the higher-derivative corrections. These new modes  can be excited during dynamical processes, such as particle scattering or gravitational collapse, even when the background solution remains identical to its GR counterpart.

In this work, we study the gravitational-wave (GW) emission by a point particle plunging into a Schwarzschild black hole within quadratic gravity and show that, although additional degrees of freedom are present and can be excited in principle, their observable impact is \emph{exponentially suppressed} in the GR limit. We interpret this suppression as the \emph{nonperturbative counterpart} of the equivalence between vacuum GR and quadratic gravity which holds order by order in the EFT expansion. This suppression ensures that 
the gravitational waveform remains essentially indistinguishable from that predicted by GR in the EFT regime, as expected~\cite{Burgess:2003jk,Endlich:2017tqa}. 

Previous studies (e.g.,~\cite{Alves:2025qcx}) have discussed observational signatures of quadratic gravity and their deviations from GR. However, these works focused on a strongly coupled regime where the EFT is not valid.
In this article, we focus instead on the weak-field regime of the full theory, when an EFT framework is in principle allowed. 
We remark that, as we argue in~\cite{Antoniou:2024jku}, it is unlikely that current astrophysical observations of black holes probe the strong-field regime of quadratic gravity (see also the discussion in Sec.\,\ref{subsec:bbsol}).

In Sec.\,\ref{sec:background} we discuss the stationary black hole background in quadratic gravity, and the equations describing a particle falling into the black hole. In Sec.\,\ref{sec:waveforms_emission} we compute the gravitational-wave emission in this process, and discuss how it approaches the GR case in the small-coupling limit. Hereafter, we adopt geometrized units in which $G=c=1$.

\section{Stationary black holes in quadratic gravity and their perturbations}
\label{sec:background}
\subsection{Action and field equations}

In four spacetime dimensions, the most general extension of Einstein gravity that includes only the Einstein--Hilbert term and quadratic curvature invariants can be written as~\cite{Stelle:1976gc,Stelle:1977ry,Whitt:1985ki}
\begin{equation}
    \mathcal{I}_0=\int {\rm d}^4 x \sqrt{-g} \left( R-\alpha C_{\mu \nu \rho \sigma} C^{\mu \nu \rho \sigma}+\beta R^2 \right)\,,
\label{eq:action}
\end{equation}
where $\sqrt{-g}$ is the determinant of the metric tensor, $R=g^{\mu\nu}R_{\mu\nu}$ is the Ricci scalar, $R_{\mu\nu}$ is the Ricci tensor, and $C_{\mu \nu \rho \sigma}$ is the Weyl tensor, such that $C_{\mu \nu \rho \sigma} C^{\mu \nu \rho \sigma}
    = 2 R_{\mu \nu} R^{\mu \nu} - \tfrac{2}{3}R^2 + \GB $, with the Gauss--Bonnet invariant being defined as $\GB \equiv R^2 - 4 R_{\mu \nu} R^{\mu \nu} + R_{\mu \nu \rho \sigma} R^{\mu \nu \rho \sigma}$.

The dimensionful couplings $\alpha$ and $\beta$ introduce additional propagating degrees of freedom: besides the usual massless spin-2 graviton, the theory predicts a massive spin-0 mode with mass squared $m^2=1/(6\beta)$ and a massive spin-2 mode with mass squared $\mu^2=1/(2\alpha)$.
We supplement the vacuum theory with the usual (minimally coupled) matter action, such that the full theory reads
\begin{equation}
   \mathcal{I}=\frac{1}{16\pi}\, \mathcal{I}_0[g_{\mu\nu}]+\mathcal{I}_m [g_{\mu\nu},\psi]\, ,
\end{equation}
where $\mathcal{I}_m=\int {\rm d}^4 x \sqrt{-g} \,\mathcal{L}_m$ is the matter action depending on the metric and (schematically) on the matter fields $\psi$.

The massive spin-2 excitation carries a kinetic term with the wrong sign, i.e. it is a ghost-like degree of freedom. This signals a violation of unitarity if quadratic gravity is regarded as a fundamental quantum theory (though alternative interpretations have been proposed~\cite{Donoghue:2021cza}). 
From an EFT perspective, however, this pathology is not problematic: as long as the energy scale is below the cut-off $\sim \alpha^{-1/2} \sim \mu$, the theory remains predictive. 
In this framework, genuine deviations from GR are not expected to appear since the corresponding corrections in the action can be removed order by order through suitable field redefinitions, at least in vacuum, see\,\cite{Burgess:2003jk,Endlich:2017tqa} and Sec.\,\ref{subsec:redef} below. 

Introducing an auxiliary tensor field $f_{\mu\nu}$ and a scalar one $\phi$, the vacuum action is equivalent to~\cite{Held:2022abx}
\begin{equation}
\begin{split}
    \mathcal{I}_0=
    \int {\rm d}^4 x \sqrt{-g} \bigg[&
    R + 2 f_{\mu\nu}G^{\mu\nu}
    +{\mu^2}\left(f_{\mu\nu}f^{\mu\nu}-f^2\right)\\
    &+\frac{3}{2}\phi\,\Box\phi-\frac{3}{2}m^2 e^{-2\phi}(e^\phi-1)^2\bigg]\,.
\end{split}
\label{eq:Lagrangian}
\end{equation}
The equations of motion for $g_{\mu\nu}$ can be found by taking variations of the action with respect to $f_{\mu\nu}$: 
\begin{equation}
     \mathcal{E}^{(g)}_{\mu\nu}\equiv G_{\mu\nu}+ \mu^2 \left(f_{\mu\nu}-f g_{\mu\nu}\right)=0
\label{eq:field_g}
\end{equation}
from which we deduce
\begin{equation}
    R_{\mu\nu} = -\mu^2 \left(f_{\mu\nu}+\frac{1}{2} f g_{\mu\nu}\right)\quad, \quad R=-3\,\mu^2 f\,.
\label{eq:Rbg}
\end{equation}
The equations of motion for $f_{\mu\nu}$ read
\begin{equation}
\begin{split}
    &
    G_{\mu\nu}+G_{(\mu}{}^\rho f_{\nu)\rho}-g_{\mu\nu}G^{\rho\sigma}f_{\rho\sigma}-fR_{\mu\nu}
    \\
    &
    +R f_{\mu\nu}+\Box f_{\mu\nu}+(\nabla_\mu\nabla_\nu-g_{\mu\nu})f
    -2\nabla_\rho\nabla_{(\mu}f_{\nu)}{}^\rho\\
    &
    +g_{\mu\nu}\nabla_\rho\nabla_\sigma f^{\rho\sigma}+\mu^2\big[\big(f^2-f_{\rho\sigma}f^{\rho\sigma}\big)/2\\
    &
    +2\big(f_\mu{}^\rho f_{\nu\rho}-f\,f_{\mu\nu}\big)\big]-8\pi\, T_{\mu\nu}=0
\end{split}
\end{equation}
where $T_{\mu\nu}\equiv -(2/\sqrt{-g})\,\delta (\sqrt{-g}\mathcal{L}_m) /\delta g^{\mu\nu}$.
Furthermore, by taking the covariant derivative of Eq.~\eqref{eq:field_g}, we get
\begin{equation}
    \mathcal{E}^{(\nabla)}_\mu\equiv\nabla^\nu f_{\mu\nu}-\nabla_\mu f=0\,.
\label{eq:nabla_f}
\end{equation}
After applying the commutation of covariant derivatives, we can re-write the equation of motion for $f_{\mu\nu}$ as follows:
\begin{equation}
\begin{split}
    \mathcal{E}^{(f)}_{\mu\nu}
    \equiv &\mu^2\left[ f_{\mu\nu}(f-1)+g_{\mu\nu}\left(f+\frac{1}{2}f^{\alpha\beta}f_{\alpha\beta}\right)\right]\\
    &+\Box f_{\mu\nu}-\nabla_\mu\nabla_\nu f+2R_{\rho\mu\sigma\nu}f^{\rho\sigma}\\
    &-\frac{3}{4}\phi(\Box-2\nabla_\mu\nabla_\nu)\phi+\frac{3}{4}m^2 e^{-2\phi}(e^\phi-1)^2\\
    &-8\pi T_{\mu\nu}=0\,.
\end{split}
\label{eq:field_f}
\end{equation}
Finally, by taking variations with respect to the scalar field we derive the scalar field equation of motion:
\begin{equation}
    \mathcal{E}^{(\phi)}
    \equiv \Box\phi+m^2e^{-2\phi}(e^\phi-1)^2=0\,.
\end{equation}
Since the scalar field equation decouples from the rest, we will not study it in the context of this work, and we will focus instead on the coupled spin-2 modes, which are sourced by matter fields.

\subsection{Black hole solutions}\label{subsec:bbsol}
Stationary, asymptotically flat solutions of quadratic gravity have vanishing Ricci scalar~\cite{Nelson:2010ig,Lu:2015cqa}, and hence also $f=0$.
There exist two branches of stationary black hole solutions: (i) Ricci-tensor flat solutions, for which ${R}_{\mu\nu}=0=f_{\mu\nu}=0$, yielding the standard GR equations in vacuum, so that the black hole solution is simply the Kerr metric -- or, in the static case, the Schwarzschild metric; (ii) Ricci-scalar flat solutions, for which ${R}=0$ but $R_{\mu\nu}\neq0$, yielding black holes endowed with nontrivial profiles of the extra gravitational degrees of freedom of the theory\,\cite{Lu:2015cqa,Lu:2015psa} (see also~\cite{Kokkotas:2017zwt}). We shall refer to these solutions as {\it hairy black holes}. For simplicity, in the following we shall only consider static, spherically symmetric black holes.

The horizon radius $r_h$ of a hairy black hole is constrained both from above and from below. 
If $r_h$ exceeds the upper bound, the corresponding solutions acquire negative mass and are therefore unphysical. 
Indeed, in contrast to the Schwarzschild solution which exists for arbitrary horizon radii and always has a monotonic mass--radius relation, the mass of the hairy black hole is a \emph{decreasing} function of the horizon radius. 
On the other hand, solutions with $r_h$ smaller than the lower bound are unstable under radial perturbations~\cite{Held:2022abx,Held:2023aap}. 
One finds that static, radially stable hairy black holes with positive mass exist only within the window $0.876 \lesssim p  \lesssim 1.143$, where the dimensionless parameter $p$ is the horizon radius in units of the coupling constant, i.e. of the massive spin-2 field:
\begin{equation}
    \label{eq:defp}
    p=\frac{r_h}{\sqrt{2\alpha}}=r_h\mu\,.
\end{equation}
The existence of an upper cutoff in $p$ indicates that the hairy black hole branch is intrinsically non-perturbative and cannot be captured within the EFT framework. 

The monopolar instability of the hairy black hole adds to the same instability occurring in the Schwarzschild branch in the same range, $p\lesssim 0.876$~\cite{Babichev:2013una,Brito:2013wya} (see also~\cite{Myung:2023ygn,Myung:2013doa}).

Finally, hairy black holes exist only below a critical mass, $M_c \simeq 0.438/\mu$\,\cite{Lu:2015cqa}. As pointed out in~\cite{Antoniou:2024jku}, for $M$ not too close to $M_c$, both the horizon and the ISCO radii of hairy black holes are significantly larger than their GR counterparts; thus, unless their mass $M$ is fine tuned to be close to the maximal value $M_c$ (which is fixed by the coupling constant), hairy black holes are likely to be incompatible with electromagnetic and GW observations.

For these reasons, in this work we will consider perturbations of the  Schwarzschild solution, 
in the regime of the theory where this solution is stable, i.e. $\sqrt{\alpha}\lesssim 1.6 M$ or, equivalently, $\mu M\gtrsim 0.44$. We will still consider the theory as exact, in order to explore the approach to the perturbative, EFT regime from a non-perturbative point of view.

\subsection{Equivalence with general relativity in an effective field theory framework}\label{subsec:redef}
As mentioned in the Introduction, within an EFT framework quadratic gravity is equivalent to GR, at least in vacuum. 
Indeed, since the quadratic interaction vanishes on-shell in GR, a first-order correction can be removed --~modulo higher-order terms~-- with a field redefinition.
In other words, the quadratic terms in the action are {\it redundant}~\cite{Burgess:2003jk}. This is a general feature of EFTs: an EFT term vanishing on-shell at the leading order can be rearranged as a term proportional to the leading-order field equation, say $\sim G^{\mu\nu}Y_{\mu\nu}$, by defining $g_{\mu\nu}\to g_{\mu\nu}+Y_{\mu\nu}$.
The corresponding change in the action removes the interaction at first order in the EFT expansion, yielding higher-order terms. Then, following the same procedure, these terms can be removed order by order with appropriate field redefinitions.
In the case of quadratic gravity, the field redefinition
\begin{equation}
    g_{\mu\nu}\to g_{\mu\nu}+\alpha\left(R_{\mu\nu}-\frac{1}{3}g_{\mu\nu}R\right)
\end{equation}
removes the Weyl term $\alpha C_{\mu\nu\rho\sigma}C^{\mu\nu\rho\sigma}=-2\alpha R_{\mu\nu}R^{\mu\nu}+\frac{2}{3}\alpha R^2$ to $O(\alpha)$.

The perturbative equivalence of quadratic gravity with GR in vacuum can also be seen by looking at the perturbative solutions around the GR vacuum background, i.e. the Ricci-flat black hole solution.
Since at the zeroth order $R_{\mu\nu}=0=R$, it is easy to see that no corrections are generated at any higher order. This does not apply, of course, to the perturbations of a hairy black hole background, which is non-perturbative in the coupling.

\subsection{Spherical-harmonic decomposition of the perturbations}
\label{sec:perturbations}
In the following  we shall study linear perturbations of the spacetime induced by a point particle.

\subsubsection{Metric field decomposition}
We introduce the following expansion for the metric and the auxiliary field:
\begin{align}
g_{\mu\nu}&=\bar{g}_{\mu\nu}+\varepsilon\, \delta g_{\mu\nu}\,,\\
f_{\mu\nu}&=\bar{f}_{\mu\nu}+\varepsilon\, \delta f_{\mu\nu}\,.
\end{align}
where $\varepsilon$ is a bookkeeping parameter, $\bar{g}_{\mu\nu}$ and $\bar{f}_{\mu\nu}$ are the background metric and auxiliary field corresponding to a static, spherically symmetric BH solution,
\begin{align}
    \bar{g}_{\mu\nu}&={\rm diag}(A(r),1/B(r),r^2,r^2\sin^2\theta)\,,\\
    \bar{f}_{\mu\nu}&={\rm diag}(C(r),1/D(r),r^2,r^2\sin^2\theta)\,,
\end{align}
and $\delta g_{\mu\nu}$, $\delta f_{\mu\nu}$ are their respective perturbations.
Hereafter, any background quantity will be denoted with a bar. 

At first order in the perturbations, Eq.~\eqref{eq:field_g} gives: 
\begin{equation}
    \delta\mathcal{E}^{(g)}_{\mu\nu}\equiv \delta G_{\mu\nu} + \mu^2 \left(\delta f_{\mu\nu}-\bar{g}_{\mu\nu}\,\delta f\right)=0
    \;,
\label{eq:pert_g}
\end{equation}
while Eq.\,\eqref{eq:field_f} gives:
\begin{equation}
\begin{split}
    \delta\mathcal{E}^{(f)}_{\mu\nu}
    &\equiv\;\bar{\Box}\delta f_{\mu\nu}+2\bar{R}_{\rho\mu\sigma\nu}\delta f^{\rho\sigma}
    -\mu^2 \delta f_{\mu\nu}\\
    &-8\pi\left(T_{\mu\nu}-\frac{1}{3}g_{\mu\nu}T +\frac{1}{3\mu^2}\bar{\nabla}_\mu\bar{\nabla}_\mu T\right)=0\,.
\end{split}
\label{eq:pert_f}
\end{equation}
Two more equations arise from the constraint~\eqref{eq:nabla_f} and the trace of~\eqref{eq:field_f}
\begin{align}
    \delta\mathcal{E}^{(\nabla)}_{\mu}\equiv \bar{\nabla}^\nu \delta f_{\nu\mu}-\frac{8\pi}{3\mu^2}\bar{\nabla}_\mu T=0\\
    \delta f-\frac{8\pi}{3\mu^2} T=0\,.
\end{align}
Note that, in vacuum, the equation for $\delta f_{\mu\nu}$ at the linear level is equivalent to Fierz-Pauli massive gravity~\cite{Fierz:1939ix}, and hence also to the linearized version of nonlinear massive gravity and bimetric theories~\cite{Brito:2013wya}. However, the coupling to matter and nonlinear interactions are different, as we shall discuss later.

We also stress that for a vanishing background scalar, i.e. $\bar{\phi}=0$, the massive scalar perturbations completely decouple from the other perturbations. They are only characterized by the massive scalar equation $\delta \mathcal{E}^{(\phi)}$ which we shall not consider in this study.

We expand the $g_{\mu\nu}$ and $f_{\mu\nu}$ tensors in spherical harmonics. We choose the Regge-Wheeler gauge~\cite{Regge:1957td} for the massless ($g_{\mu\nu}$) sector. Since this choice of gauge does not leave any residual freedom, for the decomposition of $\delta f_{\mu\nu}$ we have to consider a generic decomposition in spin-0, spin-1, and spin-2 spherical harmonics. 
The harmonic expansions of the axial and polar sectors are given below in terms of the standard scalar spherical harmonics $Y_{\ell m}$:
\begin{widetext}
\begin{equation}
    \delta g^{\ell m}_{\rm ax}=
    \begin{pmatrix}
    0 & 0 & -h_0^{\ell m} \csc\theta\,\partial_\varphi & h_0^{\ell m} \sin\theta\,\partial_\theta \\
    0 & 0 & -h_1^{\ell m} \csc\theta\,\partial_\varphi & h_1^{\ell m} \sin\theta\,\partial_\theta  \\
    * & * & 0 & 0\\
    * & * & 0 & 0
    \end{pmatrix}
    Y_{\ell m} \quad , \quad
    \delta g^{\ell m}_{\rm pol}=
    \begin{pmatrix}
    A H_0^{lm}  & H_1^{\ell m}  & 0 & 0 \\
    * &B^{-1} H_2^{\ell m} & 0 & 0 \\
    0 & 0 & r^2 H^{\ell m}  & 0 \\
    0 & 0 & 0 & r^2\sin^2\theta H^{\ell m} 
    \end{pmatrix}Y_{\ell m}\,,
\label{eq:pert_ansatz_g}
\end{equation}
\begin{equation}
\begin{split}
    &\delta f^{\ell m}_{\rm ax}=
    \begin{pmatrix}
    0 & 0 & f_0^{\ell m}\csc\theta\,\partial_{\phi} & -f_0^{\ell m}\sin\theta\,\partial_{\theta}\\
    0 & 0 & f_1^{\ell m}\csc\theta\,\partial_{\phi} & -f_1^{\ell m}\sin\theta\,\partial_{\theta} \\
    * & * & -f_2^{\ell m}\csc\theta\,X_{\ell m} & f_2^{\ell m}\sin\theta\, W_{\ell m}  \\
    * & * & * & f_2^{\ell m}\sin\theta\, X_{\ell m}
    \end{pmatrix}
    Y_{\ell m}\quad,\quad
    \begin{matrix}
        X_{\ell m}=&\;2\partial_\theta\partial_\varphi Y_{\ell m}-2\cot\theta\,\partial_\varphi Y_{\ell m}\quad\quad\\[2mm]
        W_{\ell m}=&\;\partial_\theta^2 Y_{\ell m}-\cot\theta\,\partial_\theta Y_{\ell m}-\csc^2\theta\,\partial_\varphi^2 Y_{\ell m}
    \end{matrix}~,\\[5mm]
    &\delta f^{\ell m}_{\rm pol}=
    \begin{pmatrix}
    A F_0^{lm} Y_{\ell m} & F_1^{\ell m} Y_{\ell m} & \eta_0^{\ell m}\partial_{\theta}Y_{\ell m}& \eta_0^{\ell m}\partial_{\phi}Y_{\ell m} \\
    * &B^{-1} F_2^{\ell m}Y_{\ell m} & \eta^{\ell m}_1\partial_{\theta}Y_{\ell m}& \eta_1^{\ell m}\partial_{\phi}Y_{\ell m} \\
    * & * & r^2[K^{\ell m} Y_{\ell m} +G^{\ell m} W_{\ell m}] & r^2  G^{\ell m} X_{\ell m} \\
    * & * & * & r^2\sin^2\theta[K^{\ell m} Y_{\ell m}-G^{\ell m}W_{\ell m}]
    \end{pmatrix}\,.
\end{split}\label{eq:df_exp}
\end{equation}
\end{widetext}

In the following, we will adopt this decomposition of the fields, together with a similar decomposition of the point-particle source, to study the linear response of Schwarzschild black holes in this theory. The background quantities then read $A(r)=B(r)\equiv F(r)=1-2M/r$ and $C(r)=D(r)=0$.

\subsubsection{Matter field decomposition}
For simplicity, we shall consider a massive point particle in radial infall toward the black hole. The stress-energy tensor contains only polar perturbations and can be decomposed as (see, e.g.,~\cite{Davis:1971gg,Sago:2002fe,Blazquez-Salcedo:2016enn})
\begin{equation}
    T_{\mu\nu}=
    \begin{pmatrix}
        \mathcal{A}^{(0)} & \mathcal{A}^{(1)} & 0 & 0 \\
        \mathcal{A}^{(1)} & \mathcal{A} & 0 & 0 \\
        0 & 0 & 0 & 0 \\
        0 & 0 & 0 & 0
    \end{pmatrix}
    Y_{\ell m}
\end{equation}
where
\begin{align}
    {\mathcal A}^{(0)}_{\ell m }=&m_pu^0\left(\frac{dr_p}{dt}\right)^{-1}\frac{F(r)^2}{r^2}e^{i\omega t_p}Y_{\ell m }^\star(\theta_p,\varphi_p)\, , \\ 
    {\mathcal A}^{(1)}_{\ell m }=&\sqrt{2}i m_pu^0 \frac{e^{i\omega t_p}}{r^2}Y_{\ell m }^\star(\theta_p,\varphi_p)\, ,\\
    {\mathcal A}_{\ell m }=&m_pu^0\frac{dr_p}{dt}\frac{e^{i\omega t_p}}{r^2F(r)^2}Y_{\ell m }^\star(\theta_p,\varphi_p)\, ,
\end{align}
with the particle velocity time-component being $u^0=dt_p/d\tau=\gamma F(r)^{-1}$, whereas $dr_p/dt=-F(r) \gamma^{-1}\sqrt{\gamma^2-F(r)}$, and $t_p=t_p(r)$ can be found by solving the following differential equation
\begin{equation}
    \frac{dt_p}{dr}=\left(\frac{dr_p}{dt}\right)^{-1}-\frac{\gamma F(r)^{-1}}{\sqrt{\gamma^2-F(r)}}\,,\label{eq:dtdr}
\end{equation}
with $\gamma$ being the particle's energy per unit mass (i.e., the Lorentz boost).
Note that this source excites only polar perturbations of $g_{\mu\nu}$ and $f_{\mu\nu}$
and we will focus on those in the following.
For simplicity, we shall also focus our numerical analysis on the ultrarelativistic case, $\gamma\to\infty$.

\subsection{Master equations}
To begin with, we derive the linearized equations in vacuum. As previously mentioned, these equations are equivalent to those of massive gravity~\cite{Fierz:1939ix}, which have been studied in~\cite{Brito:2013wya} in the context of black hole perturbations. They are
\begin{align}
    &\bar{\Box}\delta f_{\mu\nu}+2\bar{R}_{\rho\mu\sigma\nu}\delta f^{\rho\sigma}
    -\mu^2 \delta f_{\mu\nu}=0\,,\label{eq:pert_f_QNMs}\\[2mm]
    \begin{split}
        & \bar{\nabla}_\alpha \bar{\nabla}_\mu \delta g_{\nu\alpha} + \bar{\nabla}_\alpha \bar{\nabla}_\nu \delta g_{\mu\alpha} - \bar{\Box} \delta g_{\mu\nu} - \bar{\nabla}_\nu \bar{\nabla}_\mu \delta g \\
        &-(\bar{g}_{\mu\nu}+\delta g_{\mu\nu})(\bar{\nabla}_\alpha \bar{\nabla}^\beta \delta g_{\beta\alpha}-\bar{\Box} \delta g)\\
        &+ 2\mu^2 \left(\delta f_{\mu\nu}-\bar{g}_{\mu\nu}\,\delta f\right)=0\, .\label{eq:pert_g_QNMs}
    \end{split}
\end{align}
It is worth noting that the massive sector perturbations, $\delta f_{\mu\nu}$, are decoupled from the massless sector perturbations, $\delta g_{\mu\nu}$, since Eq.\,\eqref{eq:pert_f_QNMs} does not depend on $\delta g_{\mu\nu}$. The equation for the massless perturbations, Eq.\,\eqref{eq:pert_g_QNMs}, depends instead on the massive perturbations $\delta f_{\mu\nu}$.

By replacing the spherical harmonic decomposition discussed in Sec.\,\ref{sec:perturbations}, and by performing a Fourier transform of the perturbations to the frequency domain,
\begin{align}
    \delta f_{\mu\nu}(t,\vec{x})=\frac{1}{\sqrt{2\pi}}\int d\omega\; \delta \hat{f}_{\mu\nu}(\omega,\vec{x})e^{-i\omega t}\\
    \delta g_{\mu\nu}(t,\vec{x})=\frac{1}{\sqrt{2\pi}}\int d\omega\; \delta \hat{g}_{\mu\nu}(\omega,\vec{x})e^{-i\omega t}\,,\label{eq:fourierdec}
\end{align}
the perturbation equations with polar parity can be cast as a system of second-order ordinary differential equations, in the general form
\begin{equation}
    \frac{d^2}{dr^2}\boldsymbol{\Psi}_{\ell m}+\boldsymbol{P}_{\ell }\frac{d}{dr}\boldsymbol{\Psi}_{\ell m}+\boldsymbol{V}_{\ell }\boldsymbol{\Psi}_{\ell m}=0\,,
    \label{eq:master_equations}
\end{equation}
where $\boldsymbol{\Psi}_{1m}$ is a vector of $N$ perturbation functions, with $N=1$ for $\ell=0$, $N=2$ for $\ell=1$ and $N=4$ for $\ell\ge2$. In the following we shall discuss these cases separately. $\boldsymbol{P}_\ell$, $\boldsymbol{V}_\ell$ are $N\times N$ matrices of coefficients. Since the background is spherically symmetric, such coefficients are only functions of the radial coordinate, and do not depend on the harmonic index $m$. 

In Appendix~\ref{app:QNMs} we compute the polar parity QNMs of a Schwarzschild BH in quadratic gravity, by solving Eqs.\,\eqref{eq:master_equations} as an eigenvalue problem (note that the axial parity QNMs have been computed in Ref.\,\cite{Antoniou:2024jku}). In the next sections, we shall consider the excitation of the perturbations $\delta g_{\mu\nu}$, $\delta f_{\mu\nu}$ by a physical source, i.e. a particle radially falling into the black hole.

The equations with axial parity can be cast in a similar form. 
Since the radially falling particle does not excite axial parity perturbations, we will not discuss them here.

\subsubsection*{Vacuum equation for the monopole ($\ell=0$)}
Monopolar massless spin-2 perturbations do not propagate as GWs. Thus, we can solve for the massive perturbations alone, which are decoupled from the massless one.
The perturbation equations can be cast as a single equation for the perturbation function $K$. Thus, in this case the perturbation vector is just ${\Psi}_{00}=K^{00}$, and 
\begin{align}
P_0&=\frac{2}{r FP}\left(4 \mu ^2 r F^2+\mu ^4 r^2 (r-M)+4 \omega^2 (2 r-3 M)\right)
\label{eq:Pl0}\\
V_0&=\frac{1}{P F^2r^3}\left(-F (\mu ^6 r^6+6 \mu ^2 r^4 \omega ^2) \right.\nonumber\\
&+4 \omega ^2 \left(-12 M^2+6 M r+r^4 \omega ^2\right)\nonumber\\
    &\left.+\mu ^4 r^2 \left(r^4 \omega ^2-4 \left(8 M^2-6 M r+r^2\right)\right)\right)\,,
\label{eq:Vl0}
\end{align}
where we defined
%
$P(r)=r\left(2 \mu ^2 F+\mu ^4 r^2+4 \omega ^2\right)$.

\subsubsection*{Vacuum equations for the dipole ($\ell=1$) with polar parity}
Dipolar massless spin-2 perturbations do not propagate GWs either, thus we can solve for the massive perturbations alone. The perturbation equations can be cast as
a system of two equations involving the perturbation functions $K$ and $\eta_1$ only, i.e. Eq.\,\eqref{eq:master_equations} with $\boldsymbol{\Psi}^\top_{1m}=(K^{1m},\eta_1^{1m})$. The explicit expressions of the $2\times 2$ matrices $\boldsymbol{P}_1$, $\boldsymbol{V}_1$ are shown in Appendix\,\ref{app:coeffs}.

\subsubsection*{Vacuum equations for $\ell\geq2$ with polar parity}
In this case, the massless spin-2 sector also has dynamical degrees of freedom, which are coupled to those of the massive spin-2 sector. The perturbation equations can be cast in the form\,\eqref{eq:master_equations}, with $\boldsymbol{\Psi}^\top_{\ell m}=(K^{\ell m},\eta_1^{\ell m},G^{\ell m},H^{\ell m})$. 
Here $H^{\ell m}$ fully describes the polar sector of the massless spin-2 mode, and satisfies a second-order equation that can be solved for once the perturbations of $f_{\mu\nu}$ are known. The explicit expressions of the $4\times 4$ matrices $\boldsymbol{P}$, $\boldsymbol{V}$ are shown in Appendix\,\ref{app:coeffs}.

\subsection{Solutions with a source}
When including a source term, the above equations can be written as a system of inhomogeneous ordinary differential equations, i.e. 
\begin{equation}
    \left(\frac{d^2}{dr^2}+\boldsymbol{P}_\ell\frac{d}{dr}+\boldsymbol{V}_\ell\right)\boldsymbol{\Psi}_{\ell m}=\boldsymbol{S}_{\ell}\label{eq:master_S}
\end{equation}
where the left-hand side coincides with Eq.\,\eqref{eq:master_equations}, with $\boldsymbol{\Psi}_{\ell m}=K^{\ell m}$ for $\ell=0$, $ \boldsymbol{\Psi}^\top_{\ell m}=(K^{\ell m},\eta_1^{\ell m})$ for $\ell=1$ and $\boldsymbol{\Psi}^\top_{\ell m}=(K^{\ell m},\eta_1^{\ell m},G^{\ell m},H^{\ell m})$ for $\ell\ge2$.

In the monopolar case, the source term is  $\boldsymbol{S}_0= S^{(K)}_{0}$:
\begin{equation}
    S_0^{(K)}=-\frac{8 \gamma \sqrt{\pi } F \left(\mu ^2 r+2 i \omega \right) e^{i \omega  t_p}}{Q_0}\,.\label{eq:S0}
\end{equation}
In the dipolar case,  $\boldsymbol{S}^\top_{1 }= (S^{(K)}_{1},S^{(\eta_1)}_{1})$:
\begin{align}
    S_1^{(K)}=&\frac{F(r)}{r}\, S_1^{(\eta_{1})}\, ,\\
    S_1^{(\eta_{1})}=&\frac{8 \sqrt{3 \pi } \gamma  \left(\mu ^2 r^2+2 i r w+2\right)}{Q_{1}} e^{i \omega t_p} \, ,\label{eq:S1}
\end{align}
Finally, for $\ell\ge2$,  $\boldsymbol{S}^\top_{\ell }= (S^{(K)}_{\ell},S^{(\eta_1)}_{\ell},S^{(G)}_{\ell},S^{(H)}_{\ell})$:
\begin{align}
    S_\ell^{(K)}=&\frac{F}{r}\, S_\ell^{(\eta_{1})}\, ,\\
    S_\ell^{(\eta_1)}=&
    \frac{8 \gamma  \sqrt{\pi  (2\ell +1) } \left(\mu ^2 r^2+2 i r \omega+\ell ^2+\ell \right)}{Q_{\ell}} e^{i \omega t_p} \, ,\\
    S_\ell^{(G)}=& \frac{2 F \sqrt{\pi  (2\ell +1) } }{3 \gamma  \mu ^2 r^4}e^{i \omega t_p}\,.\label{eq:Sge2}
\end{align}
In Eqs.\,\eqref{eq:S0}-\eqref{eq:Sge2} we have defined
\begin{equation}
\begin{split}
    Q_{\ell}= & \; 4 \mu ^2 M r^2-4 M \ell  (\ell +1)-\mu ^4 r^5\\
    &-2 r^3 \left(2 \omega^2+\mu ^2 \left(\ell ^2+\ell +1\right)\right)\\
    &-r \ell  \left(\ell ^3+2 \ell ^2-\ell -2\right)\, .
\end{split}
\end{equation}

To solve the system of inhomogeneous differential equations we will employ the Green's matrix approach~\cite{Sisman:2009mk} which we summarize in Appendix~\ref{app:methods}. 
We define the ingoing/outgoing solutions of the homogeneous equations, characterized by purely ingoing/outgoing boundary conditions at the horizon/infinity, respectively.
By making use of the generalized matrix Green function, the solution is given by
\begin{equation}
    \boldsymbol{\Psi}_{\ell m}(t,r_*)=\boldsymbol{\hat{a}}\,\boldsymbol{W}_{\ell m}(r_*)\,\int dr_*'\, \underline{\boldsymbol{\mathcal{I}}}\,\boldsymbol{W}_{\ell m}^{-1}(r_*')\,\boldsymbol{\hat{b}}\boldsymbol{S}_{\ell }(r_*')
\end{equation}
where $\boldsymbol{W}_{\ell m}$ is the Wronskian matrix obtained from two independent ingoing/outgoing solutions and their first derivatives. The explicit expression of the Wronskian matrix, together with the matrices $\boldsymbol{\hat{a}}$, $\boldsymbol{\hat{b}}$, $\boldsymbol{\mathcal{I}}$, are given in Appendix~\ref{app:methods}. 

\subsection{GW stress-energy tensor}
In order to compute the GW energy flux emitted by perturbations in quadratic gravity, we must construct a suitable stress-energy tensor (SET) associated with the radiative degrees of freedom. We use the Noether current approach, in which the stress-energy tensor is derived as the conserved current associated to diffeomorphism invariance\,\cite{Saffer:2017ywl}. In order to get a symmetric tensor, we apply a generalized Belinfante procedure, adding an additional, conserved term which is covariant under $ISO(3,1)$ rather than under general coordinate transformation\,\cite{Bak:1993us,Guarrera:2007tu}. We remark that this approach requires a background Minkowski metric $\eta_{\mu\nu}$, and thus it can only be applied in quasi-Cartesian, asymptotically flat coordinates at large distances from the source. The result becomes gauge-invariant once Brill-Hartle average is performed\,\cite{Saffer:2017ywl}.

We expand the action to quadratic order in the perturbations
\begin{equation}
    S=S^{(0)}+\epsilon S^{(1)}+\epsilon^2 S^{(2)}+\mathcal{O}(\epsilon^3)\,.
    \label{eq:action_012}
\end{equation}
where $S^{(0)}$ corresponds to the background action, $S^{(1)}$ vanishes upon short-wavelength averaging\,\cite{Stein:2010pn,Saffer:2017ywl}, and $S^{(2)}$ encodes the effective GW dynamics. Explicitly
\begin{align}
     S^{(2)}=&\frac{1}{16\pi}\int  {\rm d}^4 x \sqrt{-\bar{g}}  \bigg[\bar{g}^{\mu\nu}R^{(2)}_{\mu\nu} - \delta g^{\mu\nu}R^{(1)}_{\mu\nu}\nonumber\\
     & + \frac{1}{2} \delta g\,R^{(1)} + 2 \delta f^{\mu\nu}G^{(1)}_{\mu\nu}+ 2\bar{f}^{\mu\nu}G^{(2)}_{\mu\nu}\nonumber\\
     &+{\mu^2}\left(\delta f_{\mu\nu}\delta f^{\mu\nu}-\delta f^2\right)\bigg]
     =\int d^4x\sqrt{-g}\mathcal{L}^{(2)}\,.\label{eq:S2}
\end{align}
The Lagrangian $\mathcal{L}^{(2)}$ can be written as the sum of the Einstein-Hilbert (EH) second-order action and its quadratic gravity correction (Q): 
$    \mathcal{L}^{(2)}=\mathcal{L}^{(2)}_{\rm EH}+\mathcal{L}^{(2)}_{\rm Q}$.
By considering perturbations of a Minkowski background, they read:
\begin{align}
     \mathcal{L}^{(2)}_{\rm EH}
     =&-\frac{1}{64\pi}\left[\partial_\lambda \delta g_{\alpha \beta} \partial^\lambda \delta g^{\alpha \beta} - \partial_\lambda \delta g \partial^\lambda \delta g\right.\nonumber\\
    &\left.+ 2 \partial_\lambda \delta g^{\lambda \nu} \partial_\nu \delta g - 2 \partial_\lambda \delta g^{\lambda \nu} \partial_\rho \delta g^\rho_{\ \nu} \right]\label{eq:L2EH}\\     
\mathcal{L}^{(2)}_{\rm Q}=  & \frac{\delta f^{\mu\nu}}{32\pi}\left[
-2\partial_\mu\partial_\nu\delta g^\alpha_{~\alpha}
+2\partial_\mu\partial_\alpha\delta g^\alpha_{~\nu}
+2\partial_\alpha\partial_\nu\delta g^\alpha_{~\mu}\right.\nonumber\\
&-2\partial_\alpha\partial^\alpha\delta g_{\mu\nu}+\frac{1}{2}\eta_{\mu\nu}\left(\partial^\alpha\partial_\alpha\delta g^\rho_{~\rho}
-2\partial_\alpha\partial_\beta\delta g^{\alpha\beta}\right.\nonumber\\
&\left.\partial_\alpha\partial^\alpha\delta g^\beta_{~\beta}\right)
\left.+\frac{\mu^2}{16\pi}\left(\delta f_{\mu\nu}\delta f^{\mu\nu}-\delta f^2\right)\right]\,.
\label{eq:L2Q}
\end{align}
The  conserved current corresponding to diffeomorphism invariance is
\begin{equation}
T^{\rm C}_{\mu\nu}= -\frac{\partial \mathcal{L}^{(2)}}{\partial (\delta g^{\rho\sigma,\mu})} \delta g^{\rho\sigma}{}_{,\nu}
+ \eta_{\mu\nu} \mathcal{L}^{(2)}\,,
\end{equation}
which is not symmetric. To obtain a symmetric tensor, which can be interpreted as the stress-energy tensor associated to the perturbations, we follow the Belinfante procedure\,\cite{Bak:1993us,Guarrera:2007tu} which consists in adding a suitable conserved term, which can be expressed in terms of the spin matrices\,
\begin{equation}
    \Sigma_{\nu \lambda}^{\rho\sigma}[\delta g] = \big( \eta_{\lambda \kappa} \delta_\nu^\rho - \eta_{\nu \kappa} \delta_\lambda^\rho \big) \delta g^{\kappa \sigma} + \big( \eta_{\lambda \kappa} \delta_\nu^\sigma - \eta_{\nu \kappa} \delta_\lambda^\sigma \big) \delta g^{\kappa \rho}\,.
\end{equation}
The result of this procedure is:
\begin{equation}
\begin{split}
    &T^{{\rm B}}_{\mu\nu} =  -\frac{\partial \mathcal{L}^{(2)}}{\partial (\delta g^{\rho\sigma,(\mu})} \delta g^{\rho\sigma}{}_{,\nu)}
    \\
    &-\partial_\lambda\left( \frac{\partial \mathcal{L}^{(2)}}{\partial (\delta g^{\rho\sigma,(\mu})} \Sigma^{\rho\sigma}_{\nu)\lambda} [\delta g]\right)+ \eta_{\mu\nu} \mathcal{L}^{(2)}\,.
\end{split}
\label{eq:TGW}
\end{equation}
In the wave zone, the metric and the auxiliary field perturbations describe the GWs emitted by the system. Their stress-energy tensor is the Brill-Hartle average (i.e., average over several wavelengths) of $T^{\rm B}_{\mu\nu}$. 
By focusing on the $(tr)$ component which will be relevant for the calculation of the flux, the last term inside the brackets vanishes, and we are left with 
\footnote{In massive gravity~\cite{Cardoso:2018zhm} the GW stress-energy tensor is derived from the massive gravity quadratic Lagrangian and is given by
\begin{align}
    T^{\text{GW(m)}}_{tr}=\langle &(\partial_t\delta f_{\alpha\beta})(\partial_r\delta f^{\alpha\beta})\rangle
\label{eq:TGW_massive}
\end{align}}
\begin{align}
    T^{\text{GW}}_{tr} =& -\bigg\langle 
    \frac{\partial \mathcal{L}^{(2)}}{\partial(\delta g^{\rho\sigma,(t})} \delta g^{\rho\sigma}_{~~,r)}
    +\partial_\lambda\left( \frac{\partial \mathcal{L}^{(2)}}{\partial (\delta g^{\rho\sigma,(t})} \Sigma^{\rho\sigma}_{r)\lambda} [\delta g]\right)
    \bigg\rangle
\label{eq:TGW_tr}
\end{align}
Substituting Eqs.\,\eqref{eq:L2EH},~\eqref{eq:L2Q} in Eq.\,\eqref{eq:TGW_tr} we find the  expression of $T^{\text{GW}}_{tr}$ in terms of the metric perturbations, which we explicitly show in Appendix~\ref{app:SET}.

\subsection{Energy flux}
The emitted energy flux is:
\begin{equation}
    \frac{dE}{dt}
    =-\int T^{\text{GW}}_{tr}r^2 d\Omega\,.
\label{eq:eqLGW}
\end{equation}
The perturbation functions are decomposed in the frequency domain according to Eqs.\,\eqref{eq:fourierdec}.
By applying Parseval's theorem, $\int dt \,g(t)h^*(t)=\int d\omega \tilde g(\omega)\tilde h(\omega)$ to 
\begin{equation}
    E=\int \frac{dE}{dt}dt=\int\frac{dE}{d\omega}d\omega
\end{equation}
we obtain the expression of $dE/d\omega$, by replacing in each term of the stress-energy tensor $T^{\text{GW}}_{tr}$ in Eq.\,\eqref{eq:eqLGW} the perturbations $\delta g_{\mu\nu}$, $\delta f_{\mu\nu}$ with the corresponding Fourier transforms.
For instance, $(\partial_t\delta g_{\alpha\beta})(\partial_r\delta g^{\alpha\beta})$ $\to$ $(-i\omega\delta\tilde g_{\alpha\beta})(\partial_r\delta{\tilde g}^{\alpha\beta\,*})$. Finally, the perturbations are expanded in tensor spherical harmonics as discussed in Sec.\,\ref{sec:perturbations}. Since the background is spherically symmetric, the angular integration in Eq.\,\eqref{eq:eqLGW}, using the orthogonality relations of tensor spherical harmonics, leads to an expression in the form 
\begin{equation}
    \frac{dE}{d\omega}=\sum_{\ell m}C^{\ell m}_{tr} (\omega,\mu)\,.\label{eq:fluxC}
\end{equation}
We explicitly computed $C^{\ell m}_{tr}$ only in the monopolar, $\ell=0$ case, in which only massive degrees of freedom are present at infinity. We remark that the $\ell\ge2$ case is much more involved, due to its complicated asymptotic behavior, where massless and massive modes are mixed. For instance, for $H^{\ell m}(r)$ we have for $r_*\to-\infty$, 
\begin{align}
    & H^{\ell m}_{\text{out}} \to e^{i\omega r_*} + A^{\ell m} e^{-i\omega r_*}\nonumber\\
    & H^{\ell m}_{\text{in}} \to e^{-i\omega r_*}\,,\label{eq:asymptoci_H}
\end{align}
while for $r_*\to+\infty$,
\begin{align}
    & H^{\ell m}_{\text{out}}\to \, c^{\ell m}_\omega \,e^{i{k} r_*}+e^{i \omega r_*}\\
    & H^{\ell m}_{\text{in}} \to \, (c^{\ell m}_\omega \,e^{i{k} r_*}+e^{i \omega r_*})+B (c^{\ell m}_{d-\omega} \, e^{-i{k} r_*}+e^{-i \omega r_*})\,,\nonumber \label{eq:asymptoci_inf0}
\end{align}
where we define $k\equiv \sqrt{\omega^2-\mu^2}$.
In the $\ell=0$ case, instead, the computation of $C^{\ell m}_{tr}$ is straightforward. By making use of~\eqref{eq:pert_g}-\eqref{eq:pert_f}, we find that (see Appendix\,\ref{app:SET} for details)
\begin{align}
\begin{split}
    &(F_0,F_1,F_2,K)= \; r^{M(\mu^2-2\omega^2)/(ik)}e^{i k r}\sum_{n=0}^\infty \frac{\mathcal{A}_n}{r^{n+1}}\, ,\\
    &(H_1,H_2) \sim \; r^{M(\mu^2-2\omega^2)/(ik)}e^{i k r}\sum_{n=0}^\infty \frac{\mathcal{B}_n}{r^{n}}\,.
\end{split}
\end{align}
where $\mathcal{A}_n$ and $\mathcal{B}_n$ are the $n$-order coefficients for each one of the perturbation functions, i.e. $\mathcal{A}_n=(F_{0(n)},F_{1(n)},F_{2(n)},K_{(n)})$ and $\mathcal{B}_n=(H_{1(n)},H_{2(n)})$ (the explicit expressions of these coefficients are given in Appendix~\ref{app:SET}).
At infinity the source terms do not contribute and we find that $F_0^\text{(out)}\sim-({2k^2}/{\mu^2})K^\text{(out)},\;F_1^\text{(out)}\sim({2\omega k}/{\mu^2})K^\text{(out)},\;F_2^\text{(out)}\sim-2\left[({k^2}/{\mu^2})+1\right]K,\;H_2\sim-2ik K^\text{(out)}$. By replacing in Eq.\,\eqref{eq:TGW_tr} we finally find:
\begin{equation}
\label{eq:Ctr}
    C^{00}_{tr}(\omega,\mu)=-\frac{8\omega \left(\omega^2-\mu ^2\right)^{3/2}}{\mu ^2}\vert K^{00} \vert^2\,,
\end{equation}
where the spectrum $dE/d\omega$ has been defined to be one-sided, i.e. with $\omega>0$.

For $\ell\geq1$, we compute only the field amplitudes since, as argued below, they are sufficient to capture the exponential suppression of the GR deviations.

\section{Nonperturbative suppression in emitted signals}
\label{sec:waveforms_emission}

\begin{figure}[t]
    \centering
    \includegraphics[width=0.95\linewidth]{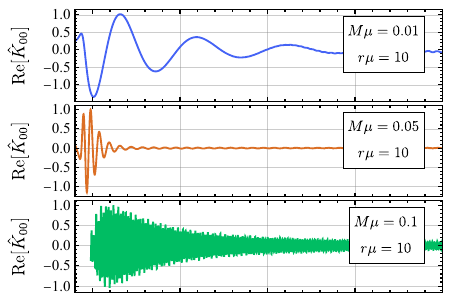}\\
    \includegraphics[width=0.95\linewidth]{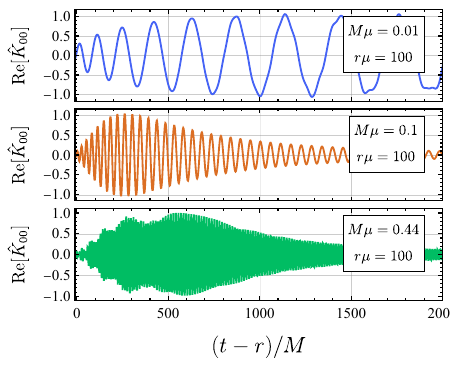}
    \caption{Normalized waveforms for the monopolar perturbations in the highly relativistic limit ($\gamma\gg 1$), for an extraction radius of $r\mu=(10,100)$ and $M\mu=(0.01,0.05,0.1)$. In each panel the amplitude has been rescaled to its maximum value.    
    }
    \label{fig:waveforms_l0}
\end{figure}

\subsection{Monopolar radiation}

We first examine the monopolar case by considering Eq.\,\eqref{eq:master_S} for $\ell=0$ in the presence of a radially infalling particle. 
In Fig.~\ref{fig:waveforms_l0} we show the waveform for a few different choices of the mass $\mu$, i.e. $M\mu=(0.01,\,0.1,\,0.44)$, which were chosen as representative of different regimes. In particular, we choose the value $M\mu=0.44$ as this lies right next to the threshold line determining the monopolar instability (as well as the bifurcation value for the hairy branch of solutions).
These waveforms are computed  using the methodology presented above to obtain the signal in the frequency domain and then performing an inverse Fourier transform.
Notice that the top three panels in Fig.~\ref{fig:waveforms_l0} correspond to an extraction radius of $r\mu=10$, while the bottom three ones are computed at $r\mu=100$, where dispersion effects are more relevant.

We then computed  the energy flux when $\mu M\gg1$, given by Eqs.\,\eqref{eq:fluxC},~\eqref{eq:Ctr}.
As expected, since the scalar degree of freedom is massive, emission at infinity is possible only for frequencies $\omega>\mu$.

\begin{figure}
    \centering
    \includegraphics[width=0.95\linewidth]{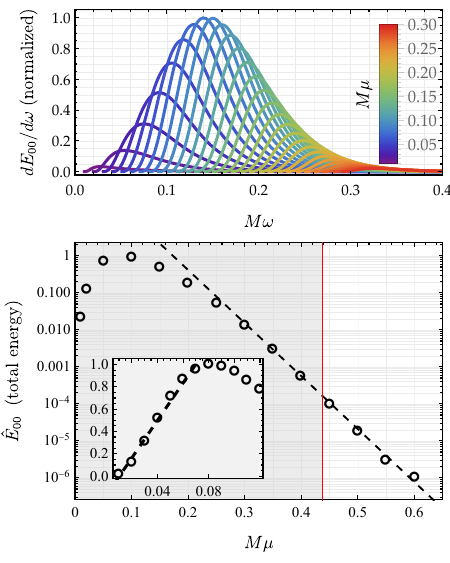}
    \caption{Top: Monopolar energy flux emitted by a relativistic radially infalling particle, for $M\mu\in[0.001,0.3]$. Each curve is normalized to its maximum value. 
    Bottom: Total emitted energy (normalized by its maximum value) as a function of $M\mu$.
    The total flux vanishes exponentially in the weak-coupling regime (i.e., as $M\mu$ increases): the dashed line corresponds to the fit $c_1 e^{-c_2 \mu M}$ with $c_1\approx 303$ and $c_2\approx -33$.
    The shaded area denotes the strong-coupling region where black hole solutions are radially unstable. For completeness, in the inset we show the behavior in the strong-coupling regime (i.e., $M\mu\to0$).} 
    \label{fig:flux_monopole}
\end{figure}

In Fig.~\ref{fig:flux_monopole} we show our main results for the monopolar flux. 
The energy spectrum $dE/d\omega$ has support only at $\omega>\mu$, peaks near the left boundary and decays to zero as $\omega\gg\mu$.

The total emitted energy can be obtained by integrating, 
\begin{equation}
    \hat E_{lm}=\int_{0}^{+\infty}d\omega \frac{dE_{lm}}{d\omega}\,.
\end{equation}
As shown in the bottom panel of Fig.~\ref{fig:flux_monopole}, the total monopole energy flux has a maximum around $\mu M\approx 0.08$ and decays linearly to zero in the \emph{strong-coupling} regime ($\mu M\to0$, $\alpha\to\infty$, see inset), in agreement with the analysis of~\cite{Cardoso:2018zhm} in massive gravity. However, both these features occur in the regime where black holes in quadratic gravity are unstable (gray shaded area in the lower panels). Interestingly, in the allowed regime $M\mu\gtrsim 0.44$ the behavior is radically different: we find that the total flux is \emph{exponentially suppressed} and well fitted by 
\begin{equation}
\label{eq:expdecay}
    \hat E_{00}\sim c_1 e^{-c_2 \mu M} \sim c_1 e^{-c_2/\sqrt{\alpha}}\,.
\end{equation}
This finding is noteworthy for several reasons. First, it demonstrates that deviations from GR are exponentially suppressed precisely in the most physically relevant and theoretically motivated regime. Second, as evident from Eq.\,\eqref{eq:expdecay}, the effect is intrinsically nonperturbative and therefore inaccessible within an EFT expansion: the Taylor series in $\alpha$ vanishes identically at every order. It can thus be viewed as a genuinely nonperturbative manifestation of the vacuum equivalence between quadratic gravity and GR within the EFT framework~\cite{Burgess:2003jk,Endlich:2017tqa}.  
As we shall show, this feature appears to be generic, extending to higher multipoles.

\subsection{Dipolar and quadrupolar radiation}

\begin{figure*}
    \centering
    \includegraphics[width=0.95\linewidth]{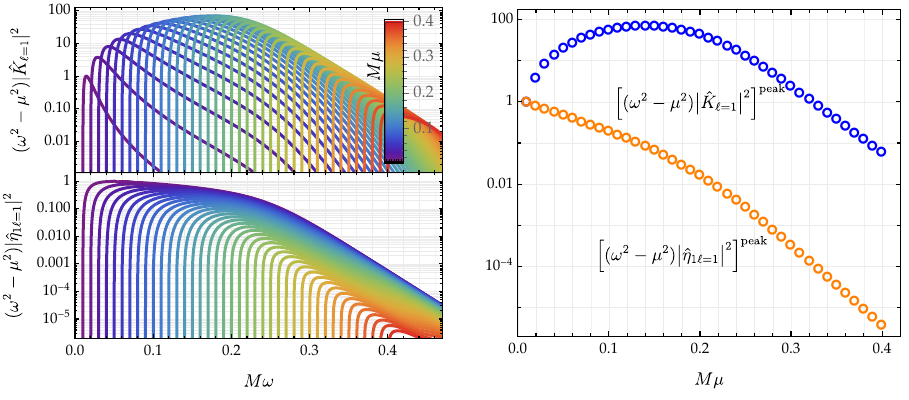}
    \caption{\textit{Left:} Amplitudes of the massive perturbation functions for $\ell=1$ for $M\mu=[0.01,0.4]$ as a function of the frequency $\omega$, normalized by their peak value for $M\mu=0.01$.
    \textit{Right:} Peak values of the amplitudes as a function of the mass $\mu$. In both panels, the log-linear scale makes the exponential suppression evident.}
    \label{fig:peak_scaling_l1}
\end{figure*}

\begin{figure*}
    \centering
    \includegraphics[width=0.95\linewidth]{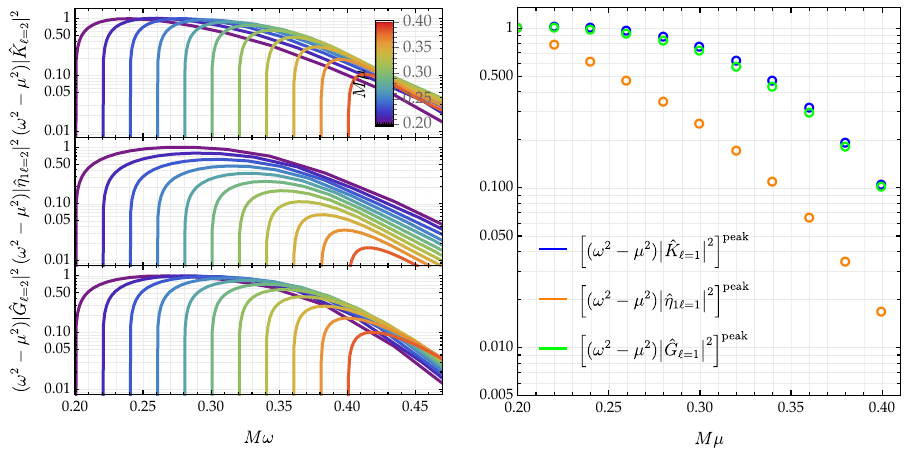}
    \caption{Same  as Fig.~\ref{fig:peak_scaling_l1} but for $\ell=2$.
    }
\label{fig:peak_scaling_l2}
\end{figure*}

We solve Eqs.~\eqref{eq:master_equations} for $\ell=1$ and $\ell=2$, using the method detailed in Appendix~\ref{app:methods}. While we are mostly interested in exploring the suppression of beyond-GR corrections when approaching the perturbative regime, for completeness we have also computed some representative examples of dipolar and quadrupolar waveforms (see Appendix~\ref{app:waveforms}).

The flux formula for $\ell>0$ is significantly more involved than in the monopolar case, and we do not derive it here. In particular, while the $\ell=1$ flux receives contributions only from the massive modes, the $\ell\geq 2$ flux also includes a component from the massless modes.
However, for massive modes, any nonanalytic dependence on $\alpha$ can arise only through the amplitudes $|K|$ and $|\eta|$ (in the dipole case), and through $|K|$, $|\eta|$, and $|G|$ (for multipoles with $\ell \geq 2$), all evaluated at infinity, as they enter the flux formula.
Likewise, in the massless case, any nonanalytic behavior can only originate from the amplitudes $|H_0|$, $|H_2|$, and $|H|$ evaluated at infinity.

Thus, in order to check any possible exponential suppression we can focus on the amplitudes.
We present this analysis in Figs.~\ref{fig:peak_scaling_l1} and~\ref{fig:peak_scaling_l2} for the massive $\ell=1$ and $\ell=2$ modes, respectively.
In the left panels, we show the amplitudes --~multiplied by a factor $(\omega^2-\mu^2)$, which is also expected in the flux formula~-- as functions of the frequency $\omega>\mu$, for various values of $\mu$. 
The behavior is very similar to the top panel of Fig.~\ref{fig:flux_monopole}: the amplitudes display a maximum whose frequency grows with $\mu$, and then decay exponentially.
To better visualize the exponential behavior, in the right panels we show the peak of the amplitude as a function of $M\mu$. 
Since the total energy involves an integral of the amplitude over frequency, the exponential suppression of the peak amplitude translates into an exponential suppression of the total energy flux, in agreement with the bottom panel of Fig.~\ref{fig:flux_monopole} in the monopolar case.

Overall, in all cases we observe an \emph{exponential suppression} as $M\mu$ increases, i.e. when approaching the perturbative regime.

Due to this strong suppression and the fact that large $M\mu$ corresponds to high-frequency radiation, exploring the regime where $M\mu\gg1$ is numerically challenging. However, as clear from Fig.~\ref{fig:flux_monopole} already when $M\mu\gtrsim 0.3$ the behavior is exponential and we did not find any evidence of a change of slope in our curves. We therefore expect that this trend would continue also for larger values of $M\mu$ and for any $\ell$.

\section{Conclusions}\label{sec:conclusions}
In this work we have investigated the dynamics of black hole perturbations in quadratic gravity, a well-motivated extension of GR obtained by augmenting the Einstein--Hilbert action with quadratic curvature terms. 
Although the theory is equivalent to vacuum GR within an EFT framework, it also admits nonperturbative features that can modify the dynamics of compact objects. 

Black holes in quadratic gravity exhibit a richer linear response than in GR, supporting additional scalar, vector, and tensor QNMs that can in principle be excited in astrophysical processes, even when the stationary background coincides with the Schwarzschild solution. 
To explore these effects, we computed the QNM spectrum of a Schwarzschild black hole and studied the gravitational-wave emission from a point particle in geodesic motion around it. 

Our analysis confirms previous results showing that the QNM spectrum is significantly modified by deviations from the standard GR modes and by the presence of additional families of massive modes. 
However, we have also demonstrated that all observable non-GR effects are exponentially suppressed in the weak-coupling limit, thereby providing a concrete nonperturbative realization of the GR-EFT equivalence anticipated by theoretical arguments. 
This finding reinforces the view that, while quadratic gravity introduces new gravitational degrees of freedom, their phenomenological impact on astrophysical black holes is strongly suppressed, making the two theories effectively indistinguishable in the weakly coupled regime.

More broadly, our results illustrate how higher-derivative corrections, though important at a fundamental level and in the ultraviolet regime, can become dynamically negligible in astrophysical scenarios dominated by low-curvature, low-frequency processes. 
This nonperturbative manifestation of the GR–EFT equivalence strengthens the theoretical foundations of quadratic gravity as a consistent EFT extension of GR and clarifies expectations for its observational signatures in the context of black hole physics.

Although our calculations focused on point particles in radial infall, the exponential suppression we uncovered appears generic and likely independent of the details of the source. 
It will be important to verify whether this behavior persists for other observables and in more general dynamical settings. 
Furthermore, it is natural to ask whether similar nonperturbative effects occur in other higher-curvature theories of gravity. 

Finally, we stress that the equivalence between quadratic gravity and GR holds strictly in vacuum and relies on the fact that quadratic curvature corrections vanish for GR black holes. 
This property no longer applies in the presence of matter, for instance in neutron-star spacetimes. 
In such systems, deviations from GR may arise already at leading order in the EFT expansion, corresponding at the nonperturbative level to deviations that vanish only polynomially in the GR limit. 
Exploring this problem is an interesting direction for future work.

\acknowledgements
We thank Gerardo Garcia Moreno
for interesting discussion. 
GA acknowledges financial support provided by FCT – Fundação para a Ciência e a Tecnologia, I.P., through the ERC-Portugal program Project ``GravNewFields''. GA also thanks the Fundação para a Ciência e Tecnologia (FCT), Portugal, for the financial support to the Center for Astrophysics and Gravitation (CENTRA/IST/ULisboa) through grant No.~\href{https://doi.org/10.54499/UID/PRR/00099/2025}{UID/PRR/00099/2025} and grant No.~\href{https://doi.org/10.54499/UID/00099/2025}{UID/00099/2025}.
This work is partially supported by the MUR FIS2 Advanced Grant ET-NOW (CUP:~B53C25001080001) and by the INFN TEONGRAV initiative.

\appendix

\section{Coefficients of the master equation for massive perturbations}\label{app:coeffs}

We here show the explicit coefficients of the master equation for the (massive) polar perturbations Eq.\,\eqref{eq:master_equations}, 
\begin{equation}
    \frac{d^2}{dr^2}\boldsymbol{\Psi}_{\ell m}+\boldsymbol{P}_{\ell }\frac{d}{dr}\boldsymbol{\Psi}_{\ell m}+\boldsymbol{V}_{\ell }\boldsymbol{\Psi}_{\ell m}=0\,,
    \label{eq:master_equations2}
\end{equation}
i.e. the coefficients of the $N\times N$ matrices $\boldsymbol{P}_\ell$, $\boldsymbol{V}_\ell$.

For $\ell=0$, $N=1$, $\boldsymbol{\Psi}=K$ and the coefficients $P_0$, $V_0$ are given in Eqs.\,\eqref{eq:Pl0}, \eqref{eq:Vl0}. 

For $\ell=1$, $N=2$, $\boldsymbol{\Psi}=(K,\eta_1)$, and the coefficients of the matrices $\boldsymbol{P}_1$, $\boldsymbol{V}_1$ are:
\begin{align}
    P_{11}= \; & 2 F \mathcal{N} r^2 (16 \mu ^2 M^2 r-M (\mu ^4 r^4+4 r^2 (5 \mu ^2\nonumber\\
    & +3 \omega ^2)-4)+r^3 (8 \mu ^2+\mu ^4 r^2+8 \omega ^2)) \\
    P_{12}= \; & 8 \mathcal{N} r F^3 \left(6 M+\mu ^2 r^3\right) & \\
    P_{21}= \; & 4 \mathcal{N} r^3 F \big[r^3 \left(\mu ^2+2 \omega ^2\right)-M \left(3 \mu ^2 r^2+2\right)\big]\\
    P_{22}= \; & 2 \mathcal{N} r F \big[-12 M^2 \left(\mu ^2 r^2+2\right)+M r (3 \mu ^4 r^4\nonumber\\
    & +2 r^2 \left(5 \mu ^2+6 \omega ^2\right)+24)+4 \mu ^2 r^4\big]\\
    V_{11}= \; & -\mathcal{N} (192 M^3+16 M^2 r (2 \mu ^4 r^4+r^2 (5 \mu ^2\nonumber\\
    &+3 \omega ^2)-10)-2 M r^2 (\mu ^6 r^6+6 \mu ^2 r^4 (3 \mu ^2+\omega ^2)\nonumber\\
    &+4 r^2 (9 \mu ^2+8 \omega ^2)-16)+r^5 (\mu ^6 r^4\nonumber\\
    &+2 \mu ^2 (r^2 \omega ^2+8)-4 \omega ^2 (r^2 \omega ^2-4)\nonumber\\
    &+\mu ^4 (10 r^2-r^4\omega ^2))) \\
    V_{12}= \; & 4 F^2 \mathcal{N}(24 M^2-2 M r (\mu ^2 r^2+6)\nonumber\\
    &+\mu ^2 r^4 (\mu ^2 r^2+4))\big/ {r} \\
    V_{21}= \; &-2 \mathcal{N} r (\mu ^2 r^2+2) (-12 M^2+M (4 r-3 \mu ^2 r^3)\nonumber\\
    &+r^4 (\mu ^2+2 \omega ^2))& \\
    V_{22}= \; &+\mathcal{N} \big[M^3 (80 \mu ^2 r+{32}/{r})-4 M^2 (7 \mu ^4 r^4\nonumber\\
    &+r^2 (46 \mu ^2+20 \omega ^2)+16)+2 M r (\mu ^6 r^6\nonumber\\
    &+2 \mu ^2 r^4 \left(10 \mu ^2+\omega ^2\right)+4 r^2 (17 \mu
    ^2+9 \omega ^2)+16)\nonumber\\
    &+r^4 (\mu ^4 r^2 (r^2 \omega ^2-12)-r^4\mu ^6+2 \mu ^2 \left(r^2 \omega ^2-14\right)\nonumber\\
    &+4 \omega ^2 \left(r^2 \omega ^2-2\right))\big]=0\, ,\\
    \frac{1}{\mathcal{N}}= \; & M \left(8 r^3-4 \mu ^2 r^5\right)+r^6 \left(6 \mu ^2+\mu ^4 r^2+4 \omega^2\right)
\end{align}

For $\ell\ge2$, $N=4$, $\boldsymbol{\Psi}=(K,\eta_1,G,H)$, and the  matrices $\boldsymbol{P}_1$, $\boldsymbol{V}_1$ are:
\begin{align}
    \boldsymbol{P}=
    \begin{pmatrix}
        {P}_{11} & {P}_{12} & {P}_{13} & 0 \\
        {P}_{21} & {P}_{22} & {P}_{23} & 0\\
        0 & 0 & {P}_{33} & 0\\
        {P}_{41} & {P}_{42} & {P}_{43} & {P}_{44}
    \end{pmatrix}\, , \\[2mm]
    \boldsymbol{V}=
    \begin{pmatrix}
        {V}_{11} & {V}_{23} & {V}_{13} & 0\\
        {V}_{21} & {V}_{22} & {V}_{23} & 0\\
        0 & {V}_{32} & {V}_{33} & 0\\
        {V}_{41} & {V}_{42} & {V}_{43} & {V}_{44}
    \end{pmatrix} \, ,
\end{align}
(note that 
the first 3 elements of the 4th column of the matrices $\boldsymbol{P}$ and $\boldsymbol{V}$ are zero, due to  the decoupling of the massive from the massless perturbations), where
\begin{align}
    P_{11}= \; & 2 F x (16 \mu ^2 M^2 r-M (\mu ^4 r^4+2 r^2 (6 \omega^2\nonumber\\
    &+\mu ^2 (\ell ^2+\ell +8))+\ell  (\ell ^3+2 \ell ^2-3 \ell -4))+\mu ^4 r^5\nonumber\\
    &+2 r^3 (4 \omega^2+\mu ^2 (\ell ^2+\ell +2))\nonumber\\
    &+r \ell  (\ell^3+2 \ell ^2-\ell -2))\big/ r \, ,\\
    P_{12}= \; & 4 F^3 x \ell  (\ell +1) (6 M\nonumber\\
    &+r (\mu^2 r^2+\ell ^2+\ell -2))\big/r^2 \, ,\\
    P_{13}= \; & 4 F^3 x \ell  \left(\ell ^3+2 \ell ^2-\ell -2\right) \, ,\\
    P_{21}= \; & 4 F x (r^3 (\mu ^2+2 \omega ^2)-M (3 \mu ^2 r^2+\ell ^2+\ell )) \, , \\
    P_{22}= \; & -2 F x ( 12 M^2 (\mu^2 r^2 + \ell^2 + \ell) + M r (-3 \mu^4 r^4 \nonumber\\
    &- 2 r^2 (6 \omega^2 + \mu^2 (\ell^2 + \ell + 3))\nonumber\\
    &+ \ell (\ell^3 + 2 \ell^2 - 13 \ell - 14))\nonumber\\
    &- 2 r^2 \ell (\ell +1) (\mu^2 r^2 + \ell^2 + \ell - 2) ) \big/r^2  \, ,\\
    P_{23}= \; & 4 F^2 r x \ell  \left(\ell ^3+2 \ell ^2-\ell -2\right) \, , \\
    P_{31}= \; & 0 \, , \\
    P_{32}= \; & 0 \, , \\
    P_{33}= \; & -2 F (M-r)/r^2 \, , \\
    V_{11}= \; & x ( -96 M^3 \ell (\ell +1) -8 M^2 ( 4 \mu^4 r^5 + r^3 (6 \omega^2\nonumber\\
    &+ 5 \mu^2 \ell (\ell +1)) + r \ell (\ell^3 + 2 \ell^2 - 11 \ell - 12)) \nonumber\\
    &+ 2 M ( \mu^6 r^8 + 3 r^6 (2 \mu^2 \omega^2 + \mu^4 (\ell^2 + \ell + 4))\nonumber\\
    &+ r^4 (2 \omega^2 (5 \ell^2 + 5 \ell + 6) + 3 \mu^2 \ell (\ell^3 + 2 \ell^2 \nonumber\\
    &+ 5 \ell + 4)) + r^2 \ell (\ell^5 + 3 \ell^4 + 3 \ell^3 + \ell^2 - 12 \ell\nonumber\\
    &- 12)) + \mu^4 r^9 (\omega^2 - \mu^2) + r^7 (4 \omega^4 - (\mu^4 (3 \ell^2\nonumber\\
    &+ 3 \ell + 4)) + 2 \mu^2 \omega^2 (\ell^2 + \ell - 3))+ r^5 \ell (\ell +1)\nonumber\\
    & \times(\omega^2 (\ell^2 + \ell - 10) - \mu^2 (3 \ell^2+ 3 \ell + 2))\nonumber\\
    & - r^3 \ell^2 (\ell +1)^2 (\ell^2 + \ell - 2) )\big/ r^4 \, , \\
    V_{12}= \; & -2 F^2 x \ell  (\ell +1) (24 M^2-2 M r (\mu ^2 r^2-3 \ell ^2\nonumber\\
    &-3 \ell +12)+r^2 (\mu ^4 r^4+\ell ^2 (2 \mu ^2 r^2-3)\nonumber\\
    &+2 \ell  (\mu ^2 r^2-2)+\ell ^4+2 \ell ^3+4))\big/ r^4 \, , \\
    V_{13}= \; & -2 F^2 x \ell  (\ell ^3+2 \ell ^2-\ell -2) (12 M\nonumber\\
    &+r (\mu ^2 r^2+\ell ^2+\ell -4))\big/r^2 \, , \\
    V_{21}= \; & 2 x (\mu ^2 r^2+\ell ^2+\ell ) (12 M^2+M r (3 \mu ^2 r^2\nonumber\\
    &+\ell ^2+\ell -6)-(r^4 (\mu ^2+2 \omega ^2)))\big/ r^2 , ,\\
    V_{22}= \; & x ( 16 M^3 (5 \mu^2 r^2 + \ell^2 + \ell) - 4 M^2 ( 7 \mu^4 r^5 \nonumber\\
    &+ r^3 (20 \omega^2 + \mu^2 (8 \ell^2 + 8 \ell + 30)) + r \ell (\ell^3 + 2 \ell^2\nonumber\\
    &+ 7 \ell + 6))+ 2 M ( \mu^6 r^8 + r^6 (2 \mu^2 \omega^2 + \mu^4 (3 \ell^2\nonumber\\
    &+ 3 \ell + 14)) + r^4 (\mu^2 (3 \ell^4 + 6 \ell^3 + 17 \ell^2 + 14 \ell + 28)\nonumber\\
    &- 2 \omega^2 (\ell^2 + \ell - 20))+ r^2 \ell (\ell^5 + 3 \ell^4 + 3 \ell^3 + \ell^2\nonumber\\
    &+ 4 \ell + 4)) + \mu^4 r^9 (\omega^2 - \mu^2)\nonumber\\
    &+ r^7 (2 \omega^2 - 3 \mu^2) (2 \omega^2 + \mu^2 (\ell^2 + \ell + 2))\nonumber\\
    &+ r^5 (\omega^2 (\ell^4 + 2 \ell^3 + 3 \ell^2 + 2 \ell - 16)\nonumber\\
    &- \mu^2 (3 \ell^4 + 6 \ell^3 + 7 \ell^2 + 4 \ell + 8))\nonumber\\
    &- r^3 \ell^2 (\ell+1)^2 (\ell^2 + \ell - 2) )\big/ r^4 \, ,
    \\
    V_{23}= \; & 2 x (\ell^2 + \ell - 2) ( 12 M^2 (\mu^2 r^2 + \ell^2 + \ell)\nonumber\\
    &- M ( 3 \mu^4 r^5 + 2 r^3 (6 \omega^2 + \mu^2 (2 \ell^2 + 2 \ell + 5))\nonumber\\
    &+ r \ell (\ell^3 + 2 \ell^2 + 11 \ell + 10))+ \mu^4 r^6 + r^4 (4 \omega^2\nonumber\\
    & + \mu^2 (\ell^2 + \ell + 2)) + 2 r^2 \ell (\ell+1) )\big/ r^2 \, ,\\
    V_{31}= \; & 0 \, ,\\
    V_{32}= \; & 2 F^2/r^3 \, , \\
    V_{33}= \; & \omega ^2-F \left(\mu ^2 r^2+\ell ^2+\ell -2\right)\big/r^2 \, .
\end{align}
The behavior of the perturbation functions close to the horizon is
\begin{align}
    (K,G,r\eta_1/A,H)\sim & \; e^{-i\omega r}(r-2M)^{-2iM\omega}\,,
\end{align}
while at infinity:
\begin{align}
\begin{split}
    &(rK,rG,\eta_1)\sim \; r^{M(\mu^2-2\omega^2)/(ik)}e^{i k r}\, ,\\
    &H \sim \; r^{M(\mu^2-2\omega^2)/(ik)}e^{i k r}+c_{\omega}\, r^{2iM\omega}e^{i \omega r}
\end{split}
\end{align}

\section{QNMs of Schwarzschild black holes}
\label{app:QNMs}

In this appendix we present the QNMs of a Schwarzschild black hole in quadratic gravity with polar parity, which were also explored in~\cite{Brito:2013wya,OuldElHadj:2024psw}. For QNMs with axial parity in quadratic gravity, we refer the reader to\,\cite{Antoniou:2024jku}. 
\subsection{QNM frequencies and damping times}
We solve the perturbation equations\,\eqref{eq:master_equations2}, 
in the frequency domain, with boundary conditions corresponding to purely ingoing/outgoing waves at the horizon/infinity.

\begin{figure}
    \centering
    \includegraphics[width=0.95\linewidth]{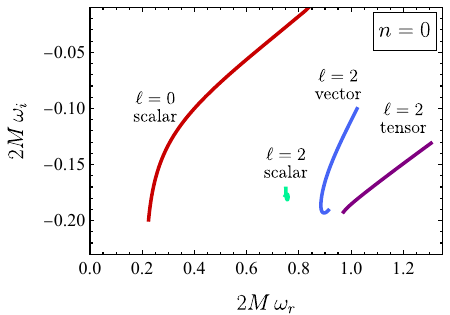}
    \caption{Fundamental ($n=0$) quasinormal modes of a Schwarzschild black hole in quadratic gravity, $\mu\in [0,0.5]$, with $\ell=0,2$, with polar parity, computed using a matrix-valued continued fraction method~\cite{Pani:2013pma}.}
    \label{fig:QNMs}
\end{figure}

Massive spin-2 perturbations exist for all values of $\ell\ge0$. Thus, we determine the QNMs for monopolar ($\ell=0$)  and quadrupolar ($\ell=2$) perturbations.
We perform the calculations by employing both a 200-step continued-fraction method and a direct integration approach (see also~\cite{Pani:2013pma,Antoniou:2024gdf}). We  present the results in Fig.~\ref{fig:QNMs}, where we show the fundamental ($n=0$) QNM complex frequency in quadratic gravity, with $\mu\in[0,0.5]$, for $\ell=0,2$.

\begin{figure*}
    \centering
    \includegraphics[width=1\linewidth]{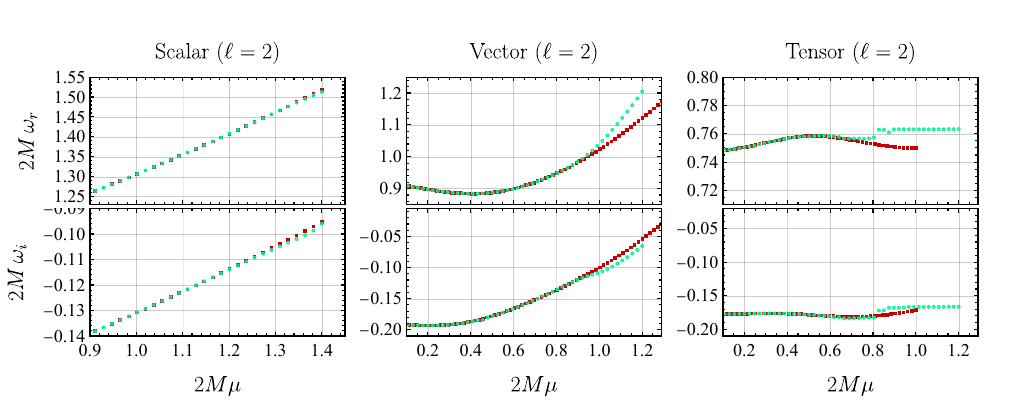}
    \caption{Scalar, vector and tensor fundamental ($n=0$) QNMs with $\ell=2$ produced with two different methods: the continued fraction method (modified Leaver's method) shown with red color and a shooting-direct integration method shown in green. The two methods agree perfectly until reaching some ``threshold'' values for $2M\mu$ (around $1.25$ for scalar, $0.95$ for vector and $0.6$ for tensor modes).}
    \label{fig:QNMs_methods}
\end{figure*}

To validate the numerical results, in  Fig.~\ref{fig:QNMs_methods} we show the  frequencies of the $\ell=2$ scalar, vector and tensor fundamental modes, as functions of $2M\mu$, computed using the continued fraction method and the direct integration approach, where the latter considers 10 terms in the boundary expansions at horizon and infinity.
The scalar, vector and tensor $\ell=2$ fundamental modes yield consistent results between the two methods below some threshold values located at $2M\mu \approx 1.25,\, 0.95,\, 0.6$ respectively.
These values lie well-within the regime in which we observe an exponential suppression of the amplitudes (see main text).

\subsection{Radial instability of monopolar perturbations}
\begin{figure}
    \centering
    \includegraphics[width=0.95\linewidth]{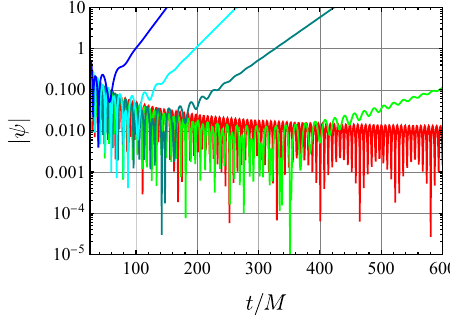}
    \caption{Time domain profile of the monopolar perturbation for $M\mu=[0.2,0.3,0.35,0.4,0.5]$ with colors ranging from blue to red respectively.}
    \label{fig:time_domain}
\end{figure}

Monopolar massive perturbations are particularly interesting,
since they are known to be subject to a long-wavelength instability~\cite{Babichev:2013una,Brito:2013wya}.
In order to study such instability, we solve the equation for monopolar perturbations
in the time domain.
We follow~\cite{Held:2022abx,Konoplya:2025afm}, according to which the monopolar perturbation can be described by a single master equation:
\begin{equation}
   \left[-\frac{\partial^2}{\partial t^2} +\frac{\partial^2 }{\partial r^2} + V(r)\right]\psi(t,r)=0\ ,
\end{equation}
where the potential $V(r)$ and the new master function $\psi(t,r)$ can be found in~\cite{Held:2022abx} (see Eqs.~(32)-(36) there).
To study the time evolution of $\psi$ we will make use of a set of lightcone coordinates $u=t-r_*$ and $v=t+r_*$, in which the previous equation can be re-written as: 
\begin{equation}
\bigg[4\,\frac{\partial^2}{\partial u\, \partial v}+V(u,v)\bigg]\psi(u,v)=0\ .
\end{equation}
As initial conditions we consider a static Gaussian pulse, \textit{i.e.},
\begin{equation}
    \psi(0,v)= e^{-\frac{v^2}{2\sigma^2}}\, , \; \psi(u,0)=0\, .
\end{equation}
We choose the width of the Gaussian pulse to be $\sigma=5M$ and ensure that our results are independent of these specific choices.
In Fig.~\ref{fig:time_domain} we show the profile of $\psi$ as a function of time normalized by the BH mass, for $M\mu=[0.2,0.3,0.35,0.4,0.5]$ and with the colors ranging from blue to red respectively.
The curves confirm the presence of an instability for $M\mu<0.43$, consistently with the results of\,\cite{Brito:2013wya,Konoplya:2025afm}.

\section{Numerical methods}\label{app:methods}

We here discuss the main numerical methods employed in this work. The reader can find further details in~\cite{Leaver:1986gd,Glampedakis:2003dn,Berti:2006wq,Pani:2013pma,Silva:2024ffz}.

\subsection{Matrix-valued direct forward integration}

The forward integration technique may be used to search for the QNMs by incorporating a shooting method. Let us consider a system of $N$ coupled differential equations.
We assume an appropriate expansion for the perturbation functions near the event horizon of the BH
\begin{equation}
\Psi^{(j)}=\;(r-r_h)^w\sum_{n=0} \psi_n^{(j)}(r-r_h)^{n+n_{(j)}}\,,
\end{equation}
where $j=1,\ldots N$ and the powers $n_{(j)}$ are specific to the particular perturbation function $\Psi^{(j)}$.
We find $N$ independent variables for the horizon boundary conditions. The remaining coefficients $\psi_n^{(j)}$ are determined in terms of the $N$ independent ones. We then perform $N$ integrations from the horizon outwards by imposing the aforementioned boundary conditions. The behavior at infinity has the following form
\begin{equation}
    \Psi^{(j)}\sim A^{(j)} e^{-i k r}r^{-\frac{q}{k}+m_{(j)}}+B^{(j)} e^{i k r}r^{\frac{q}{k}+m_{(j)}}\,,
\end{equation}
where $q$ depends on the equations at hand (e.g. on whether massive or massless modes are considered) and the powers $n_{(j)}$ are specific to the particular perturbation function $\Psi^{(j)}$.
Then, QNM solutions correspond to $B^{(j)}=0$.
We define the following matrix
\begin{equation}
\boldsymbol{S} (\omega)=
 \begin{pmatrix}
  B_{1}^{(1)} & B_{1}^{(2)} & \ldots & B_{1}^{(N)} \\
  B_{2}^{(1)} & B_{2}^{(2)} & \ldots & \ldots \\
  \ldots  & \ldots  & \ldots & \ldots  \\
  B_{N}^{(1)} & \ldots & \ldots & B_{N}^{(N)}
 \end{pmatrix}\,,
\end{equation}
each component of which, is made of the coefficients $B_i^{(j)}$. The superscript $j$ denotes a particular vector of the chosen basis, and corresponds to one of the diffferent perturbation functions.
The subscript $i$ characterizes each one of the $N$ independent integrations.
The QNM frequency $\omega_0=\omega_R+i\omega_I$ will then correspond to the solutions of
\begin{equation}
\det|\bm{S}(\omega_0)|=0\,.
\end{equation}
This is enforced by shooting for $\omega$ and minimizing the determinant at large distances.

\subsection{Matrix-valued continued fraction method}

For the \textit{continued fraction} method we make use of the following ansatz~\cite{Pani:2013pma}
\begin{equation}
    \Psi^{(i)}\sim e^{-i \omega r_*} r^{\nu} e^{-q r}\sum_n a_n^{(i)} \psi_n(r)\,\label{eq:ansatz}
\end{equation}
where $\Psi^{(i)}$ for different values of $i$ are the different perturbation functions.

For $\ell\geq 2$ the polar perturbation functions corresponding to the massive mode, satisfy a system of coupled differential equations. Substituting the ansantz\,\eqref{eq:ansatz} we find a three-term matrix-valued recurrence relation of the form
\begin{align}
&\boldsymbol{\alpha}_0 \mathbf{U}_1+\boldsymbol{\beta}_0 \mathbf{U}_0=0\,,\\
&\boldsymbol{\alpha}_1 \mathbf{U}_{2}+\boldsymbol{\beta}_1 \mathbf{U}_1+\boldsymbol{\gamma}_1 \mathbf{U}_{0}=0\,,\\
&\boldsymbol{\alpha}_2 \mathbf{U}_{3}+\boldsymbol{\beta}_2 \mathbf{U}_2+\boldsymbol{\gamma}_2 \mathbf{U}_{1}+\boldsymbol{\delta}_2 \mathbf{U}_{0}=0\,,\\
&\boldsymbol{\alpha}_n \mathbf{U}_{n+1}+\boldsymbol{\beta}_n \mathbf{U}_n+\boldsymbol{\gamma}_n \mathbf{U}_{n-1}+\boldsymbol{\delta}_n \mathbf{U}_{n-2}\nonumber\\
&\qquad\qquad\qquad\qquad+\boldsymbol{\epsilon}_n \mathbf{U}_{n-3}=0\,,\qquad n>2\,,
\end{align}
which can be reduced by 2 orders by writing:
\begin{align}
&\boldsymbol{\alpha}_0^{(2)} \mathbf{U}_1+\boldsymbol{\beta}_0^{(2)} \mathbf{U}_0=0\,,\\
&\boldsymbol{\alpha}_n^{(2)} \mathbf{U}_{n+1}+\boldsymbol{\beta}_n^{(2)} \mathbf{U}_n+\boldsymbol{\gamma}_n^{(2)} \mathbf{U}_{n-1}=0\,,\qquad n>0\,,
\end{align}
where the second order coefficients are given in terms of the first order ones by
\begin{align}
&\boldsymbol{\alpha}_n^{(2)}=\boldsymbol{\alpha}_n^{(1)},\\
&\boldsymbol{\beta}_n^{(2)}=\boldsymbol{\beta}_n^{(1)}-\boldsymbol{\delta}_n^{(1)}\left[ \boldsymbol{\gamma}_{n-1}^{(2)} \right]^{-1} \boldsymbol{\alpha}_{n-1}^{(2)},\\
&\boldsymbol{\gamma}_n^{(2)}=\boldsymbol{\gamma}_n^{(1)}-\boldsymbol{\delta}_n^{(1)}\left[\boldsymbol{\gamma}_{n-1}^{(2)}\right]^{-1} \boldsymbol{\beta}_{n-1}^{(2)},
\end{align}
and the first order ones in terms of the zeroth order coefficients by
\begin{align}
&\boldsymbol{\alpha}_n^{(1)}=\boldsymbol{\alpha}_n^{(0)},\\
&\boldsymbol{\beta}_n^{(1)}=\boldsymbol{\beta}_n^{(0)}-\boldsymbol{\epsilon}_n^{(0)}\left[ \boldsymbol{\delta}_{n-1}^{(1)} \right]^{-1} \boldsymbol{\alpha}_{n-1}^{(1)},\\
&\boldsymbol{\gamma}_n^{(1)}=\boldsymbol{\gamma}_n^{(0)}-\boldsymbol{\epsilon}_n^{(0)}\left[\boldsymbol{\delta}_{n-1}^{(1)}\right]^{-1} \boldsymbol{\beta}_{n-1}^{(1)},\\
&\boldsymbol{\delta}_n^{(1)}=\boldsymbol{\delta}_n^{(0)}-\boldsymbol{\epsilon}_n^{(0)}\left[\boldsymbol{\delta}_{n-1}^{(1)} \right]^{-1}\boldsymbol{\gamma}_{n-1}^{(1)}\, .
\end{align}
The quantity $\mathbf{U}_n=\left(a_n^{(1)},a_n^{(2)},a_n^{(3)}\right)$ is a three-dimensional vectorial coefficient and $\boldsymbol{\alpha}_n$, $\boldsymbol{\beta}_n$, $\boldsymbol{\gamma}_n$, $\boldsymbol{\delta}_n$, $\boldsymbol{\epsilon}_n$ are $3\times 3$ matrices.

The matrix-valued three-term recurrence relation can be solved using matrix-valued continued fractions. The QNM (or quasibound) frequencies are solutions to the equation $\mathbf{M}\mathbf{U}_0=0$, where
\begin{equation}
\mathbf{M}\equiv \boldsymbol{\beta}_0+\boldsymbol{\alpha}_0 \mathbf{R}^{\dagger}_0\,,
\end{equation}
with $\mathbf{U}_{n+1}=\mathbf{R}^{\dagger}_n \mathbf{U}_n$ and
\begin{equation}
\mathbf{R}^{\dagger}_n=-\left(\boldsymbol{\beta}_{n+1}+\boldsymbol{\alpha}_{n+1}\mathbf{R}^{\dagger}_{n+1}\right)^{-1} \boldsymbol{\gamma}_{n+1}\,.
\end{equation}
For nontrivial solutions we then solve numerically
\begin{equation}
\det |\mathbf{M}|=0\,.
\end{equation}

\subsection{Matrix-valued Green's function}

Let us consider a system of $N$ coupled differential equations of second order with sources, i.e.
\begin{equation}
    \boldsymbol{A}\boldsymbol{\Psi}''+\boldsymbol{A}\boldsymbol{\Psi}'+\boldsymbol{C}\boldsymbol{\Psi}=\boldsymbol{S}\,,
    \label{eq:system}
\end{equation}
where the vector $\boldsymbol{\Psi}$ is composed of the individual perturbation functions
\begin{equation}
    \boldsymbol{\Psi}=
    \begin{pmatrix}
    \Psi^{(1)} \\
    \Psi^{(2)} \\
    \vdots \\[2mm]
    \Psi^{(n)}
    \end{pmatrix}\,.
\end{equation}
From the $2N$ independent solutions of the homogeneous equation, we may construct the Wronskian of the system, i.e.
\begin{widetext}
    \begin{equation}
    \boldsymbol{W}=
    \begin{pmatrix}
    \Psi^{(1) \textrm{in}}_1 & \Psi^{(1) \textrm{in}}_2 & \ldots & \Psi^{(1) \textrm{in}}_n & \Psi^{(1) \textrm{up}}_1 & \Psi^{(1) \textrm{up}}_2 & \ldots & \Psi^{(1) \textrm{up}}_n\\
    \Psi^{(2) \textrm{in}}_1 & \Psi^{(2) \textrm{in}}_2 & \ldots & \Psi^{(2) \textrm{in}}_n & \Psi^{(2) \textrm{up}}_1 & \Psi^{(2) \textrm{up}}_2 & \ldots & \Psi^{(2) \textrm{up}}_n\\
    \vdots & \vdots & \ddots & \vdots & \vdots & \vdots & \ddots & \vdots\\[2mm]
    \Psi^{(n) \textrm{in}}_1 & \Psi^{(n) \textrm{in}}_2 & \ldots & \Psi^{(n) \textrm{in}}_n & \Psi^{(n) \textrm{up}}_1 & \Psi^{(n) \textrm{up}}_2 & \ldots & \Psi^{(n) \textrm{up}}_n\\
    \partial_{r_*}\Psi^{(1) \textrm{in}}_1 & \partial_{r_*}\Psi^{(1) \textrm{in}}_2 & \ldots & \partial_{r_*}\Psi^{(1) \textrm{in}}_n & \partial_{r_*}\Psi^{(1) \textrm{up}}_1 & \partial_{r_*}\Psi^{(1) \textrm{up}}_2 & \ldots & \partial_{r_*}\Psi^{(1) \textrm{up}}_n\\
    \partial_{r_*}\Psi^{(2) \textrm{in}}_1 & \partial_{r_*}\Psi^{(2) \textrm{in}}_2 & \ldots & \partial_{r_*}\Psi^{(2) \textrm{in}}_n & \partial_{r_*}\Psi^{(2) \textrm{up}}_1 & \partial_{r_*}\Psi^{(2) \textrm{up}}_2 & \ldots & \partial_{r_*}\Psi^{(2) \textrm{up}}_n\\
    \vdots & \vdots & \ddots & \vdots & \vdots & \vdots & \ddots & \vdots\\[2mm]
    \partial_{r_*}\Psi^{(n) \textrm{in}}_1 & \partial_{r_*}\Psi^{(n) \textrm{in}}_2 & \ldots & \partial_{r_*}\Psi^{(n) \textrm{in}}_n & \partial_{r_*}\Psi^{(n) \textrm{up}}_1 & \partial_{r_*}\Psi^{(n) \textrm{up}}_2 & \ldots & \partial_{r_*}\Psi^{(n) \textrm{up}}_n\\   
    \end{pmatrix}\,.
\end{equation}
\end{widetext}

The Green function matrix is then given by~\cite{Sisman:2009mk}
\begin{equation}
\boldsymbol{G}(r_*,r'_*)=
\left\{\,
\begin{aligned}
&-\boldsymbol{\hat{a}}\,\boldsymbol{W}(r_*)\,\overline{\boldsymbol{\mathcal{I}}}\,\boldsymbol{W}^{-1}(r_*')\,\boldsymbol{\hat{b}}, & r_*<{r}_*',\\
&\quad\, \boldsymbol{\hat{a}}\,\boldsymbol{W}(r_*)\,\underline{\boldsymbol{\mathcal{I}}}\,\boldsymbol{W}^{-1}(r_*')\,\boldsymbol{\hat{b}}, & r_*>{r}_*'.
\end{aligned}
\right.
\end{equation}
where the special matrices appearing above are given by
\begin{gather}
    \boldsymbol{\hat{a}}_{n\times 2n}=
    \begin{pmatrix}
    \boldsymbol{I} & \boldsymbol{0}
    \end{pmatrix}
    \quad,\quad
    \boldsymbol{\hat{b}}_{2n\times n}=
    \begin{pmatrix}
    \boldsymbol{0} \\
    \boldsymbol{I}
    \end{pmatrix}\,,\\[2mm]
    \overline{\boldsymbol{\mathcal{I}}}_{2n\times 2n}=
    \begin{pmatrix}
    \boldsymbol{I} & \boldsymbol{0} \\
    \boldsymbol{0} & \boldsymbol{0}
    \end{pmatrix}
    \quad,\quad
    \underline{\boldsymbol{\mathcal{I}}}_{2n\times 2n}=
    \begin{pmatrix}
    \boldsymbol{0} & \boldsymbol{0} \\
    \boldsymbol{0} & \boldsymbol{I}
    \end{pmatrix}\,,
\end{gather}
The general solution can then be expressed in terms of the Green function as:
\begin{equation}
\begin{split}
    \boldsymbol{\Psi}_{\ell m}(\omega,r_*)=&\int_{-\infty}^{+\infty}{dr'_*}\boldsymbol{G}_{\ell m}(\omega,r_*,r'_*)\boldsymbol{S}_{\ell m}(\omega,r_*)
\end{split}
\end{equation}

It is then straightforward to check that when $n=1$, $\Psi^{(1)}\equiv \phi$ the standard result for the Green function is retrieved, i.e.
\begin{equation}
\begin{split}
   G_{\omega\ell}(r_*,r'_*)=\frac{1}{W}\big[ & \Theta(r_*-r_*')\phi^{\textrm{up}}(r_*)\phi^{\textrm{in}}(r_*')\\
   & +\Theta(r_*'-r_*)\phi^{\textrm{up}}(r_*')\phi^{\textrm{in}}(r_*) \big]\, 
\end{split}
\end{equation}
where $\Theta(r_*-r_*')$ is the Heaviside step function.

\section{SET derivation}
\label{app:SET}
Here we present the full expression of the SET in Eq.~\eqref{eq:TGW_tr}:
\begin{widetext}
\begin{align}
&T_{tr}^\text{GW}=\frac{1}{64\pi}\bigg[
2 (\nabla_r \delta f^{\lambda\rho}) (\nabla_t \delta g_{\lambda\rho}^*)
- 2 g^{\lambda\rho} (\nabla_r \delta f^\sigma_\sigma) (\nabla_t \delta g_{\lambda\rho}^*)
- 2 \delta g{}_r^\rho (\nabla_t \nabla_\sigma \delta g^{\sigma *}_\rho)
+ 2 (\nabla_t \delta g_{r\rho}) (\nabla_\lambda \delta g^{\lambda\rho *})
\nonumber\\
&\qquad
+ 2 (\nabla_r \delta g_{\lambda\rho}) (\nabla_t \delta f^{\lambda\rho *})
- 2 g^{\lambda\rho} (\nabla_r \delta g_{\lambda\rho}) (\nabla_t \delta f^{\sigma *}_\sigma)
- (\nabla_r \delta g^{\lambda\rho}) (\nabla_t \delta g_{\lambda\rho}^*)
+ g^{\lambda\rho} (\nabla_r \delta g^\sigma_\sigma) (\nabla_t \delta g_{\lambda\rho}^*)
\nonumber\\
&\qquad
- 2 \delta g{}_t^\rho (\nabla_r \nabla_\sigma \delta g^{\sigma *}_\rho)
- (\nabla_r \delta g_{\lambda\rho}) (\nabla_t \delta g^{\lambda\rho *})
+ g^{\lambda\rho} (\nabla_r \delta g_{\lambda\rho}) (\nabla_t \delta g^{\sigma *}_\sigma)
+ 2 (\nabla_r \delta g_{t\rho}) (\nabla_\lambda \delta g^{\lambda\rho *})
\nonumber\\
&\qquad
- 2 (\nabla_r \delta g_{t\sigma}) (\nabla_\rho \delta g^{\rho\sigma *})
- 2 (\nabla_t \delta g_{r\sigma}) (\nabla_\rho \delta g^{\rho\sigma *})
- 2 (\nabla_t \delta g_{r}^\lambda) (\nabla_\rho \delta g{}^\rho_\lambda)^*
- 8 (\nabla^\lambda \delta f_{tr}) (\nabla_\rho \delta g{}^\rho_\lambda)^*
\nonumber\\
&\qquad
- 4 (\nabla_r \delta f^{\lambda\rho}) (\nabla_\rho \delta g_{t\lambda}^*)
+ 4 g^{\lambda\rho} (\nabla_r \delta f^\sigma_\sigma) (\nabla_\rho \delta g_{t\lambda}^*)
+ 2 (\nabla_r \delta g^{\lambda\rho}) (\nabla_\rho \delta g_{t\lambda}^*)
- 2 (\nabla_r \delta g_{t\lambda}) (\nabla_\rho \delta g{}^\rho_\lambda)^*
\nonumber\\
&\qquad
- 4 (\nabla_t \delta f^{\lambda\rho}) (\nabla_\rho \delta g_{r\lambda}^*)
+ 4 g^{\lambda\rho} (\nabla_t \delta f^\sigma_\sigma) (\nabla_\rho \delta g_{r\lambda}^*)
+ 2 (\nabla_t \delta g^{\lambda\rho}) (\nabla_\rho \delta g_{r\lambda}^*)
- 2 g^{\lambda\rho} (\nabla_t \delta g^\sigma_\sigma) (\nabla_\rho \delta g_{r\lambda}^*)
\nonumber\\
&\qquad
- 2 g^{\lambda\rho} (\nabla_r \delta g^\sigma_\sigma) (\nabla_\rho \delta g_{t\lambda}^*)
+ 4 \delta g{}_t^\lambda (\nabla_\rho \nabla_\lambda \delta f_r^{\rho *})
- 8 \delta g^{\lambda\rho} (\nabla_\rho \nabla_\lambda \delta f_{tr}^*)
+ 4 \delta g{}_r^\lambda (\nabla_\rho \nabla_\lambda \delta f_t^{\rho *})
\nonumber\\
&\qquad
- 4 \delta g{}_r^\lambda (\nabla_\rho \nabla_t \delta f^\rho_\lambda)^*
+ 4 g^{\lambda\rho} \delta g_{r\lambda} (\nabla_\rho \nabla_t \delta f^\sigma_\sigma)^*
+ 4 g_{tr} \delta g^{\lambda\rho} (\nabla_\rho \nabla_\lambda \delta f^\sigma_\sigma)^*
- 2 g_{tr} \delta g^{\lambda\rho} (\nabla_\rho \nabla_\lambda \delta g^\sigma_\sigma)^*
\nonumber\\
&\qquad
- 4 \delta g{}_t^\rho (\nabla_\rho \nabla_r \delta f^\sigma_\sigma)^*
- 2 \delta g^{\lambda\rho} (\nabla_\rho \nabla_r \delta g_{t\lambda})^*
+ 2 \delta g{}_r^\lambda (\nabla_\rho \nabla_t \delta g^\rho_\lambda)^*
- 2 g^{\lambda\rho} \delta g_{r\lambda} (\nabla_\rho \nabla_t \delta g^\sigma_\sigma)^*
\nonumber\\
&\qquad
- 4 \delta g{}_t^\lambda (\nabla_\rho \nabla_r \delta f^\rho_\lambda)^*
+ 4 g^{\lambda\rho} \delta g_{t\lambda} (\nabla_\rho \nabla_r \delta f^\sigma_\sigma)^*
- 4 \delta g{}^\rho_t (\nabla_\rho \nabla_r \delta f^\sigma_\sigma)^*
+ 2 \delta g{}_t^\lambda (\nabla_\rho \nabla_r \delta g^\rho_\lambda)^*
\nonumber\\
&\qquad
- 2 \delta g^{\lambda\rho} (\nabla_\rho \nabla_r \delta g_{t\lambda})^*
+ 2 \delta g{}_t^\lambda (\nabla_\rho \nabla_r \delta g^\rho_\lambda)^*
- 2 g^{\lambda\rho} \delta g_{t\lambda} (\nabla_\rho \nabla_r \delta g^\sigma_\sigma)^*
+ 2 \delta g{}_t^\rho (\nabla_\rho \nabla_r \delta g^\sigma_\sigma)^*
\nonumber\\
&\qquad
+ 4 \delta g{}_t^\lambda (\nabla_\rho \nabla^\rho \delta f_{r\lambda})^*
+ 4 \delta g{}_r^\lambda (\nabla_\rho \nabla^\rho \delta f_{t\lambda})^*
- 4 g^{\lambda\rho} \delta g_{r\lambda} (\nabla_\rho \nabla_\sigma \delta f^\sigma_t)^*
+ 4 \delta g{}^\rho_r (\nabla_\rho \nabla_\sigma \delta f^\sigma_t)^*
\nonumber\\
&\qquad
- 4 g^{\lambda\rho} \delta g_{t\lambda} (\nabla_\rho \nabla_\sigma \delta f^\sigma_r)^*
+ 4 \delta g{}^\rho_t (\nabla_\rho \nabla_\sigma \delta f^\sigma_r)^*
+ 2 g^{\lambda\rho} \delta g_{r\lambda} (\nabla_\rho \nabla_\sigma \delta g^\sigma_t)^*
+ 4 g_{tr} \delta g^{\lambda\rho} (\nabla_\rho \nabla_\sigma \delta g^\sigma_\lambda)^*
\nonumber\\
&\qquad
+ 2 g^{\lambda\rho} \delta g_{t\lambda} (\nabla_\rho \nabla_\sigma \delta g^\sigma_r)^*
+ 4 (\nabla_\lambda \delta g_{r\rho}) (\nabla^\rho \delta f^\lambda_t)^*
+ 2 (\nabla_r \delta g_{t\rho}) (\nabla^\rho \delta f^\lambda_\lambda)^*
+ 2 (\nabla_t \delta g_{r\rho}) (\nabla^\rho \delta f^\lambda_\lambda)^*
\nonumber\\
&\qquad
- 4 (\nabla_t \delta g_{\lambda\rho}) (\nabla^\rho \delta f^\lambda_r)^*
+ 4 (\nabla_\lambda \delta g_{t\rho}) (\nabla^\rho \delta f^\lambda_r)^*
- 4 (\nabla_r \delta g_{\lambda\rho}) (\nabla^\rho \delta f^\lambda_t)^*
+ 4 (\nabla_\rho \delta g_{r\lambda}) (\nabla^\rho \delta f^\lambda_t)^*
\nonumber\\
&\qquad
+ 4 (\nabla_\rho \delta g_{t\lambda}) (\nabla^\rho \delta f^\lambda_r)^*
- (\nabla_r \delta g_{t\rho}) (\nabla^\rho \delta g^\lambda_\lambda)^*
- (\nabla_t \delta g_{r\rho}) (\nabla^\rho \delta g^\lambda_\lambda)^*
+ 2 g^{\lambda\rho} (\nabla_r \delta g_{\lambda\rho}) (\nabla_\sigma \delta f^\sigma_t)^*
\\
&\qquad
+ 2 g^{\lambda\rho} (\nabla_t \delta g_{\lambda\rho}) (\nabla_\sigma \delta f^\sigma_r)^*
- 4 g^{\lambda\rho} (\nabla_\rho \delta g_{t\lambda}) (\nabla_\sigma \delta f^\sigma_r)^*
- 4 g^{\lambda\rho} (\nabla_\rho \delta g_{r\lambda}) (\nabla_\sigma \delta f^\sigma_t)^*
- g^{\lambda\rho} (\nabla_r \delta g_{\lambda\rho}) (\nabla_\sigma \delta g^\sigma_t)^*
\nonumber\\
&\qquad
- g^{\lambda\rho} (\nabla_t \delta g_{\lambda\rho}) (\nabla_\sigma \delta g^\sigma_r)^*
+ 2 g^{\lambda\rho} (\nabla_\rho \delta g_{t\lambda}) (\nabla_\sigma \delta g^\sigma_r)^*
+ 2 g^{\lambda\rho} (\nabla_\rho \delta g_{r\lambda}) (\nabla_\sigma \delta g^\sigma_t)^*
- 2 (\nabla_\rho \delta g^{\rho\sigma}) (\nabla_\sigma \delta g_{tr})^*
\nonumber\\
&\qquad
- 2 (\nabla_\rho \delta g^{\rho\sigma}) (\nabla_\sigma \delta g_{rt})^*
- 2 (\nabla_r \delta f^\rho_\rho) (\nabla_\sigma \delta g^\sigma_t)^*
+ (\nabla_r \delta g^\rho_\rho) (\nabla_\sigma \delta g^\sigma_t)^*
+ 4 (\nabla_r \delta f^\sigma_\rho) (\nabla_\sigma \delta g{}^\rho_t)^*
\nonumber\\
&\qquad
- 2 (\nabla_t \delta f^\rho_\rho) (\nabla_\sigma \delta g^\sigma_r)^*
+ (\nabla_t \delta g^\rho_\rho) (\nabla_\sigma \delta g{}^\sigma_r)^*
+ 4 (\nabla_\rho \delta f^\rho_t) (\nabla_\sigma \delta g{}^\sigma_r)^*
+ 4 g_{tr} (\nabla_\lambda \delta g^{\lambda\rho}) (\nabla_\sigma \delta g{}^\sigma_\rho)^*
\nonumber\\
&\qquad
+ 4 g_{tr} (\nabla^\rho \delta f^\lambda_\lambda) (\nabla_\sigma \delta g{}^\sigma_\rho)^*
- 2 g_{tr} (\nabla^\rho \delta g^\lambda_\lambda) (\nabla_\sigma \delta g{}^\sigma_\rho)^*
- 2 \delta g_{rt} (\nabla_\sigma \nabla_\rho \delta g^{\rho\sigma})^*
- 2 \delta g_{tr} (\nabla_\sigma \nabla_\rho \delta g^{\rho\sigma})^*
\nonumber\\
&\qquad
- 2 \delta g_{rt} (\nabla_\sigma \nabla^\sigma \delta f^\rho_\rho)^*
- 2 \delta g_{tr} (\nabla_\sigma \nabla^\sigma \delta f^\rho_\rho)^*
+ \delta g_{rt} (\nabla_\sigma \nabla^\sigma \delta g^\rho_\rho)^*
+ \delta g_{tr} (\nabla_\sigma \nabla^\sigma \delta g^\rho_\rho)^*
\nonumber\\
&\qquad
- 2 (\nabla_r \delta g_{t\sigma}) (\nabla^\sigma \delta f^\rho_\rho)^*
- 2 (\nabla_t \delta g_{r\sigma}) (\nabla^\sigma \delta f^\rho_\rho)^*
- 2 (\nabla_\sigma \delta g_{rt}) (\nabla^\sigma \delta f^\rho_\rho)^*
- 2 (\nabla_\sigma \delta g_{tr}) (\nabla^\sigma \delta f^\rho_\rho)^*
\nonumber\\
&\qquad
+ (\nabla_r \delta g_{t\sigma}) (\nabla^\sigma \delta g^\rho_\rho)^*
+ (\nabla_t \delta g_{r\sigma}) (\nabla^\sigma \delta g^\rho_\rho)^*
+ (\nabla_\sigma \delta g_{rt}) (\nabla^\sigma \delta g^\rho_\rho)^*
+ (\nabla_\sigma \delta g_{tr}) (\nabla^\sigma \delta g^\rho_\rho)^*
\bigg]+\text{c.c.}
\nonumber
\end{align}
where c.c. means complex conjugate.
The $(tr)$ component appearing in the integrand of~\eqref{eq:eqLGW} is given in terms of the perturbation functions below: 
\begin{equation}
\begin{aligned}
  T^{\text{GW}}_{tr}=\frac{1}{64\pi} \bigg( 
  &2 r^2 \omega^2 F_2 H_1^* - 8 i r \omega H_1 H_1^* - 4 H_1^* H_2 
  + r^2 \omega^2 H_1^* H_2 + 8 i r \omega F_2 H_2^* + 4 H_1 H_2^* \\
  &- r^2 \omega^2 H_1 H_2^* + 4 i r \omega H_2 H_2^* - 4 r^2 \omega^2 H_1^* K 
  - 12 i r \omega H_2^* K - 4 r H_1^* F_0' + 2 i r^2 \omega H_2^* F_0' \\
  &- 8 i r^2 \omega H_1^* F_1' + 4 r H_1^* F_2' + 2 i r^2 \omega H_2^* F_2' 
  - 4 i r^2 \omega H_1^* H_1' + 4 r H_2^* H_1' - 4 i r^2 \omega H_1 H_1^*{}' \\
  &- 4 r H_2 H_1^*{}' + 2 r^2 F_0' H_1^*{}' + 2 r^2 F_2' H_1^*{}' 
  - 6 r H_1^* H_2' + i r^2 \omega H_2^* H_2' + r^2 H_1^*{}' H_2' \\
  &- 2 r F_1 \left( 8 i \omega H_1^* + r \omega^2 H_2^* + 4 i r \omega H_1^*{}' - 2 H_2^*{}' \right) 
  + 2 i r^2 \omega F_2 H_2^*{}' + 2 r H_1 H_2^*{}' + i r^2 \omega H_2 H_2^*{}' \\
  &- 2 r^2 F_1' H_2^*{}' - r^2 H_1' H_2^*{}' 
  + 2 r \omega F_0 \left( r \omega H_1^* + i \left( 2 H_2^* + r H_2^*{}' \right) \right) \\
  &- 16 r H_1^* K' - 4 i r^2 \omega H_2^* K' - 4 r^2 H_1^*{}' K' 
  - 8 r^2 H_1^* K''
  \bigg) +\text{c.c.}
\end{aligned}
\end{equation}
\end{widetext}
In order to reach our final result~\eqref{eq:Ctr} for the $C_{tr}$ coefficient, we made use of the expression above along with the following relations between the coefficients for the series expansion of the perturbation functions at infinity:

\begin{figure*}
    \centering
    \includegraphics[width=0.45\linewidth]{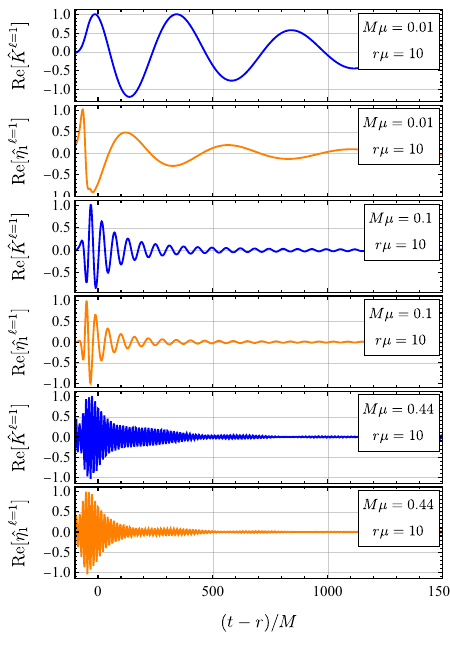}
    \includegraphics[width=0.45\linewidth]{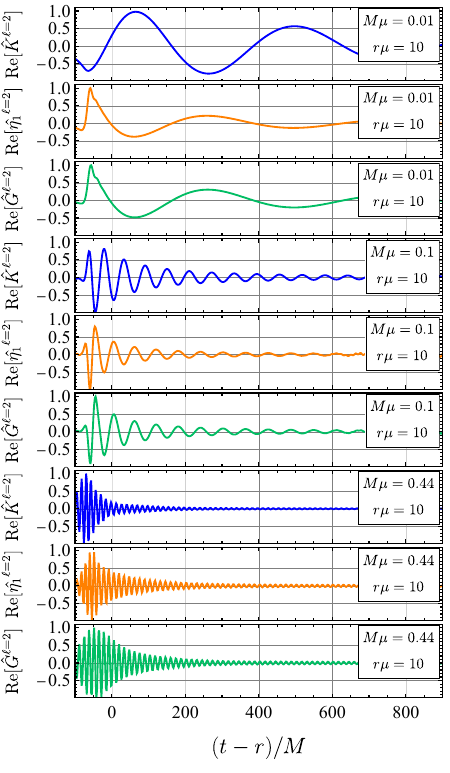}
    \caption{Normalized waveforms for the dipolar and quadrupolar perturbations in the highly relativistic limit ($\gamma\gg 1$), for an extraction radius of $r\mu=(10,100)$ and $M\mu=(0.01,0.05,0.1$).}
    \label{fig:waveforms_l1l2}
\end{figure*}
\begin{widetext}
\begin{align}
    H_{1(n=0)} &= i \omega K_{(n=0)} \, , \\
    H_{1(n=1)} &= \frac{4 \mu^4 + 5 i k^3 M \omega^2 + 3 k^4 M^2 \omega^2 - 4 \omega^4 - i k M \omega^4 - M^2 \omega^6 + 2 k^2 (\omega^2 + 3 M^2 \omega^4)}{2 k^3 \omega} K_{(n=0)}\, ,  \\
    H_{1(n=2)} &= \frac{1}{8 k^7 \omega} \bigg[ -18 k^6 M \omega^2 + 22 k^4 M \omega^4 - 144 i k^5 M^4 \omega^6 + 352 k^2 M^3 \omega^8 + 4 M^3 \omega^2 (10 \mu^8 - 61 \mu^6 \omega^2 \nonumber \\
    & \quad + 115 \mu^4 \omega^4 - 64 \omega^8) - i k \Big( 8 \mu^6 (2 + 3 M^2 \mu^2) + 3 \mu^4 (-8 - 21 M^2 \mu^2 + 3 M^4 \mu^4) \omega^2 - 17 M^2 \mu^4 \omega^4 \nonumber \\
    & \quad + (8 - 24 M^4 \mu^4) \omega^6 + 56 M^2 \omega^8 + 16 M^4 \omega^{10} \Big) + 24 i k^3 \left( 5 M^2 \omega^6 + M^4 (-3 \mu^4 \omega^4 + 4 \omega^8) \right) \bigg] K_{(n=0)}\, , \\
    H_{2(n=0)} &= -2 i k K_{(n=0)}\, , \\
    H_{2(n=1)} &= -\frac{3 i k^3 M + i k M \omega^2 + 2 k^2 (-1 + 6 M^2 \omega^2) + M^2 (3 \mu^4 - 4 \omega^4)}{k^2} K_{(n=0)} \, , \\
    H_{2(n=2)} &= \frac{M}{4 k^5} \bigg[ 144 i k^4 M^3 \omega^4 + k^3 (18 \omega^2 - 224 M^2 \omega^4) + 2 k (3 \mu^4 + 14 M^2 \mu^6 - 78 M^2 \mu^4 \omega^2 - 3 \omega^4 \nonumber \\
    & \quad + 64 M^2 \omega^6) + i M (33 \mu^6 + 9 M^2 \mu^8 - 137 \mu^4 \omega^2 - 24 M^2 \mu^4 \omega^4 + 104 \omega^6 + 16 M^2 \omega^8) \nonumber \\
    & \quad - 24 i k^2 (7 M \omega^4 + M^3 (-3 \mu^4 \omega^2 + 4 \omega^6)) \bigg] K_{(n=0)}\, ,
\end{align}

\begin{align}
    K_{(n=1)} &= \frac{i}{2 (\mu^3 - \mu \omega^2)^2} \Big( 12 k^5 M^2 \omega^2 + i M \mu^4 (-\mu^2 + \omega^2) + k^3 (3 M^2 \mu^4 + 2 \omega^2 - 16 M^2 \omega^4) \nonumber \\
    & \quad + k (-2 \mu^4 - 3 M^2 \mu^4 \omega^2 + 2 \omega^4 + 4 M^2 \omega^6) \Big) K_{(n=0)} \, , \\
    K_{(n=2)} &= \frac{1}{8 k^6 \mu^2} \bigg[ -\mu^6 (8 + 9 M^4 \mu^4) + M^2 \mu^6 (-17 + 72 M^2 \mu^2) \omega^2 + 40 k^4 M^2 \omega^4 + 96 i k^5 M^3 \omega^4 \nonumber \\
    & \quad + 41 M^2 \mu^4 \omega^4 + 8 \omega^6 - 24 M^2 \omega^8 - 64 M^4 \omega^{10} + 2 i k M ( -7 \mu^6 + \mu^4 (25 + 8 M^2 \mu^2) \omega^2  \nonumber \\
    & \quad - 42 M^2 \mu^4 \omega^4 - 18 \omega^6 + 32 M^2 \omega^8 ) + k^2 ( 17 M^2 \mu^6 - 41 M^2 \mu^4 \omega^2 + 24 (-1 + 8 M^4 \mu^4) \omega^4 + 48 M^2 \omega^6 \nonumber \\
    & \quad + 64 M^4 \omega^8 ) - 4 i k^3 \left( -9 M \omega^4 + M^3 (4 \mu^6 - 21 \mu^4 \omega^2 + 40 \omega^6) \right) \bigg] K_{(n=0)} \, , \\
    K_{(n=3)} &= \frac{M}{48 k^{10} \mu^2} \bigg[ 5520 k^4 M^2 \omega^8 - 1824 i k^5 M^3 \omega^8 + 3 k^2 \Big( -7 M^2 \mu^{10} + 210 M^2 \mu^8 \omega^2 - 1023 M^2 \mu^6 \omega^4 \nonumber \\
    & \quad + 1988 M^2 \mu^4 \omega^6 + 8 (17 + 352 M^4 \mu^4) \omega^8 - 2880 M^2 \omega^{10} + 256 M^4 \omega^{12} \Big) - 3 \Big( \mu^{10} (-16 + 39 M^4 \mu^4) \nonumber \\
    & \quad + \mu^8 (56 - 7 M^2 \mu^2 - 411 M^4 \mu^4) \omega^2+ 2 \mu^6 (-12 + 105 M^2 \mu^2 + 802 M^4 \mu^4) \omega^4 - \mu^4 (104 + 1023 M^2 \mu^2 \nonumber \\
    & \quad + 2960 M^4 \mu^4) \omega^6 + 1988 M^2 \mu^4 \omega^8 + 8 (11 + 184 M^4 \mu^4) \omega^{10} - 1168 M^2 \omega^{12} + 256 M^4 \omega^{14} \Big) \nonumber \\
    & \quad + i k M \Big( 9 \mu^{10} (-2 + 3 M^4 \mu^4) + \mu^8 (140 + 199 M^2 \mu^2 - 324 M^4 \mu^4) \omega^2 + \mu^6 (-610 - 1443 M^2 \mu^2 \nonumber \\
    & \quad + 1512 M^4 \mu^4) \omega^4 - 4 \mu^4 (-314 - 753 M^2 \mu^2 + 864 M^4 \mu^4) \omega^6 - 1608 M^2 \mu^4 \omega^8 \nonumber \\
    & \quad + 192 (-4 + 9 M^4 \mu^4) \omega^{10} - 160 M^2 \omega^{12} + 512 M^4 \omega^{14} \Big) - i k^3 \Big( -1152 M \omega^8 + 64 M^5 \omega^8 (63 \mu^4 + 8 \omega^4) \nonumber \\
    & \quad + M^3 (199 \mu^{10} - 1443 \mu^8 \omega^2 + 3012 \mu^6 \omega^4 - 1608 \mu^4 \omega^6 - 1216 \omega^{10}) \Big) \bigg] K_{(n=0)} \, , \\
    F_{0(n=0)} &= 2 \left( 1 - \frac{\omega^2}{\mu^2} \right) K_{(n=0)} \, , \\
    F_{0(n=1)} &= -\frac{M}{k \mu^2} \Big( k^3 + 3 k \omega^2 - 12 i k^2 M \omega^2 + i M (-3 \mu^4 + 4 \omega^4) \Big) K_{(n=0)} \, , \\
    F_{0(n=2)} &= \frac{M}{4 k^5 \mu^2} \bigg[ 300 i k^4 M^2 \omega^4 + 192 k^5 M^3 \omega^4 - 2 i k^2 (8 M^2 \mu^6 - 62 M^2 \mu^4 \omega^2 + 11 \omega^4 + 156 M^2 \omega^6) \nonumber \\
    & \quad + 2 i (-\mu^6 + (-4 \mu^4 + 8 M^2 \mu^6) \omega^2 - 62 M^2 \mu^4 \omega^4 + 5 \omega^6 + 54 M^2 \omega^8) + k M (5 \mu^6 + 9 \mu^4 (-5 + M^2 \mu^2) \omega^2 \nonumber \\
    & \quad - 72 M^2 \mu^4 \omega^4 + 40 \omega^6 + 64 M^2 \omega^8) - 3 k^3 (24 M \omega^4 + M^3 (3 \mu^6 - 24 \mu^4 \omega^2 + 64 \omega^6)) \bigg] K_{(n=0)} \, , \\
    F_{1(n=0)} &= \frac{2 k \omega}{\mu^2} K_{(n=0)} \\
    F_{1(n=1)} &= \frac{\omega}{k^2 \mu^2} \Big( 3 k^3 M + k M \omega^2 - 2 i k^2 (-1 + 6 M^2 \omega^2) + i M^2 (-3 \mu^4 + 4 \omega^4) \Big) K_{(n=0)} \, , \\
    F_{1(n=2)} &= -\frac{M \omega}{4 k^5 \mu^2} \bigg[ 144 k^4 M^3 \omega^4 + 2 i k^3 \omega^2 (-9 + 112 M^2 \omega^2) - 2 i k (3 \mu^4 + 14 M^2 \mu^6 - 78 M^2 \mu^4 \omega^2 \nonumber \\
    & \quad  - 3 \omega^4 + 64 M^2 \omega^6)  + M (33 \mu^6 + 9 M^2 \mu^8 - 137 \mu^4 \omega^2 - 24 M^2 \mu^4 \omega^4 + 104 \omega^6 + 16 M^2 \omega^8) \nonumber \\
    & \quad - 24 k^2 (7 M \omega^4 + M^3 (-3 \mu^4 \omega^2 + 4 \omega^6)) \bigg] K_{(n=0)} \, , \\
    F_{2(n=0)} &= -\frac{2 \omega^2}{\mu^2} K_{(n=0)} \, , \\
    F_{2(n=1)} &= \frac{1}{(\mu^3 - \mu \omega^2)^2} \Big( -9 k^2 M \omega^4 + 2 i k^3 \omega^2 (-1 + 6 M^2 \omega^2) + 5 M \omega^2 (-\mu^4 + \omega^4) \nonumber \\
    & \quad + i k (2 \mu^4 + 3 M^2 \mu^4 \omega^2 - 2 \omega^4 - 4 M^2 \omega^6) \Big) K_{(n=0)} \, , \\
    F_{2(n=2)} &= \frac{1}{4 k^6 \mu^2} \bigg[ 8 \mu^6 + 3 M^2 \mu^6 (4 + 3 M^2 \mu^2) \omega^2 - 157 k^4 M^2 \omega^4 + 204 i k^5 M^3 \omega^4 + 9 M^2 \mu^4 \omega^4 + 8 (-1 + 15 M^4 \mu^4) \omega^6 \nonumber \\
    & \quad - 21 M^2 \omega^8 - 128 M^4 \omega^{10} + 2 i k M (6 \mu^6 - 29 \mu^4 \omega^2 - 20 M^2 \mu^4 \omega^4 + 23 \omega^6 + 22 M^2 \omega^8) \nonumber \\
    & \quad + k^2 (-12 M^2 \mu^6 - 9 M^2 \mu^4 \omega^2 + 24 (1 + 3 M^4 \mu^4) \omega^4 + 82 M^2 \omega^6 + 192 M^4 \omega^8) \nonumber \\
    & \quad - 2 i k^3 (29 M \omega^4 + 4 M^3 (-5 \mu^4 \omega^2 + 19 \omega^6)) \bigg] K_{(n=0)} \, ,
\end{align}
\end{widetext}
where the index $n$ denotes the expansion order.

\section{Dipolar and quadrupolar waveforms} \label{app:waveforms}

For completeness, here we show some examples of the dipolar and quadrupolar (Fig.~\ref{fig:waveforms_l1l2}) waveforms, focusing on the massive spin-2 sector. We choose the same parameters and extraction radii as in Fig.~\ref{fig:waveforms_l0}.

\bibliography{bibnote}

\end{document}